# Comparative Framework Analysis for Enterprise Generative AI Applications: Chatbot, Automation, and Oracle-to-PostgreSQL Migration

Oleg Grynets
EPAM Systems
McLean, Virginia, USA
oleg_grynets@epam.com

Olena Pochernina
EPAM Systems
Lviv, Ukraine
olena_pochernina@epam.com

Alona Seletska
EPAM Systems
Kyiv, Ukraine
alona_seletska@epam.com

Daryna Tukalo
EPAM Systems
Kyiv, Ukraine
daryna_tukalo@epam.com

Dmytro Kostetskyi
EPAM Systems
Lviv, Ukraine
dmytro_kostetskyi@epam.com

Ivan Fedorchuk
EPAM Systems
Vinnytsia, Ukraine
ivan_fedorchuk@epam.com

Vasyl Lyashkevych
EPAM Systems
Lviv, Ukraine
vasyl_lyashkevych@epam.com

***Abstract*—This study compares framework suitability across three classes of enterprise generative AI applications: a documentation-based development assistant, an email and inquiry automation system, and an Oracle-to-PostgreSQL migration tool. The analysis evaluates component boundaries, orchestration, policy retrieval or reasoning, model integration, deterministic validation, persistence, observability, and operational efficiency. Across the three applications, the evidence supports layered architectures in which probabilistic components generate proposals, while deterministic components retain authority over routing, authorization, validation, persistence, idempotency, and final outcomes. The results indicate that framework suitability depends on the application, deployment conditions, and component responsibility: search quality, workflow control, safety behavior, and migration validation cannot be reduced to a single cross-application assessment. Therefore, the study substantiates the need for framework selection at the component level, supported by explicit contracts, application-specific evidence, and independent validation boundaries.**

***Keywords*—*generative AI, agentic systems, framework evaluation, retrieval-augmented generation, workflow orchestration, durable execution, software migration, deterministic validation***

## I. INTRODUCTION

### A. Background and Motivation

Generative artificial intelligence has expanded the range of enterprise tasks that can be supported through natural-language interpretation, retrieval, content generation, and code transformation. However, the suitability of a GenAI framework cannot be established from model capability alone. Enterprise applications also require explicit control over state, external integrations, authorization, validation, observability, and failure recovery. These requirements become especially important when generated output is used as evidence, initiates a side effect, or becomes an executable software artifact [1]–[8].

Framework selection is difficult because products described as agent, retrieval, workflow, evaluation, or observability frameworks often provide overlapping capabilities while imposing different operational assumptions. A framework that is appropriate for a synchronous conversational turn may be unsuitable for a workflow that pauses for human approval. A generic retrieval framework may add unnecessary complexity when exact rules are authoritative. Similarly, fluent generated code cannot be accepted solely because it is syntactically plausible; it requires an independent validation boundary appropriate to the target environment [9]–[23].

This article examines these issues through three enterprise application classes with materially different requirements. The first is a documentation-grounded development assistant that answers technical questions and accesses live engineering systems. The second is an automation system that classifies inbound requests, drafts responses, and controls side-effecting actions. The third is an Oracle-to-PostgreSQL migration system that analyzes source code, selects a translation strategy, validates generated PostgreSQL, and records per-unit outcomes. Together, these applications cover synchronous interaction, durable asynchronous execution, retrieval, tool use, deterministic policy, code generation, and executable validation.

The motivating architectural principle is that probabilistic generation should not implicitly own deterministic decision authority. Language models may interpret unstructured input, propose an action, draft a response, or generate a candidate translation, while deterministic components retain authority over schema enforcement, authorization, idempotency, dependency constraints, validation, and terminal outcomes. This separation does not eliminate model error, but it limits its operational consequences and makes system behavior more inspectable [1], [4], [15].

Security and evaluation studies additionally identify indirect prompt injection, judge bias, and the need for explicit retrieval- and model-evaluation procedures [24]–[27]. Recent work on GenAI-assisted software engineering further addresses architecture representation, specification determinacy, intelligent monitoring, code migration, and evolution-aware system analysis [28]–[34]. Broader studies of software functional state and LLM-based software engineering provide additional context for lifecycle-aware monitoring and code-generation risks [35]–[37].

Recent software-engineering benchmarks and execution-oriented studies provide complementary evidence for evaluating generated code, workflow behavior, testing adequacy, and reproducibility [38]–[46]. Durable-processing and transactional-workflow principles motivate explicit recovery, replay, and compensation boundaries [47], while tool-using and reflective-agent studies motivate controlled action and bounded repair mechanisms [48], [49]. Code representation, translation, structural evaluation, and reproducibility studies provide additional methodological foundations for executable software validation [50]–[58].

### B. Research Question and Objectives

The primary research question is: how well do selected GenAI frameworks and supporting technologies fit the component responsibilities, integration constraints, and operational risks of three distinct enterprise applications?

The study addresses this question through four objectives:

- identify the functional boundaries and contracts of the principal components in each application;
- compare selected frameworks and considered alternatives against consistent criteria, including capability coverage, reliability, security, integration complexity, observability, maintainability, and operational efficiency;
- distinguish implemented behavior and measured outcomes from architectural intent, untested alternatives, and proposed future capabilities; and
- derive cross-application findings about where general-purpose frameworks are sufficient, where specialized mechanisms are justified, and where deterministic controls must remain application-owned.

The evaluation does not seek a universally superior framework or a single numerical ranking across the applications. Retrieval quality, classification behavior, durable execution, and migration validity represent different outcome constructs. Comparative conclusions are therefore based on requirement fit, contract coherence, measured trade-offs, and evidence maturity rather than on direct aggregation of unlike metrics.

### C. Scope and Contributions

The scope is limited to the architecture and observed behavior of the three application prototypes and to the framework alternatives relevant to their defined components. The analysis covers orchestration, retrieval and knowledge access, model and prompt management, tool and enterprise integration, durable state, deterministic policy, validation, observability, evaluation, and bounded feedback. It does not claim production-scale performance, universal model behavior, complete coverage of enterprise security requirements, or full Oracle-to-PostgreSQL semantic equivalence.

The article makes four contributions. First, it provides a common evaluation method for comparing framework fit across applications whose outputs and execution models are not directly comparable. Second, it presents component-level analyses that separate framework capability from application-owned policy and deterministic controls. Third, it relates architectural choices to observed outcomes and limitations, including retrieval trade-offs, tool-routing errors, durability requirements, validation strength, and model-use efficiency. Fourth, it identifies recurring design conditions under which layered combinations of frameworks are more suitable than a single general-purpose agent abstraction for the evaluated application classes and component contracts [2]–[11].

The resulting contribution is architectural rather than promotional. Frameworks are assessed according to the requirements they satisfy, the complexity they introduce, and the evidence available for their behavior. Unsupported production claims and unimplemented options are retained as limitations or future work. The remainder of the article defines the evaluation method, analyzes each application separately, and then compares the resulting architectural patterns and measured trade-offs.

## II. Evaluation Scope and Method

### A. Applications and Architectural Boundaries

This article evaluates three GenAI-based enterprise applications: a documentation-grounded chatbot development assistant, an automation email and request triage system, and an Oracle-to-PostgreSQL code migration system. The applications differ in their primary outputs, execution timescales, and consequences of error. They are therefore compared through a common architectural rubric rather than through a single aggregate performance score.

The chatbot accepts developer questions and produces answers grounded in internal documentation and, when necessary, live enterprise tools. Its architectural boundary includes conversation orchestration, knowledge retrieval, LLM and prompt management, tool calling and integrations, and memory, evaluation, and observability. The main risks are unsupported answers, stale or unavailable information, uncontrolled tool use, and loss of conversational state.

The automation system accepts inbound email or request events, classifies and routes them, drafts a response, validates the draft, authorizes a proposed side effect, and either executes the action or pauses for human approval. Its boundary includes workflow and pipeline orchestration, the agent decision layer, enterprise integrations, business logic and state, and monitoring and evaluation. The main risks are incorrect routing, unauthorized action, duplicate side effects, non-recoverable pauses, and incomplete operational evidence.

The migration system accepts Oracle SQL and PL/SQL source and produces PostgreSQL artifacts together with per-unit validation outcomes and run reports. Its boundary includes source analysis and parsing, dependency planning, migration intelligence, knowledge and rules, validation and executable verification, and feedback and optimization. The main risks are incomplete source analysis, incorrect dependency ordering, invalid translation, overuse of model calls, and treating target-side validation as proof of cross-database equivalence.

The comparison follows the principle that probabilistic components propose content or actions, while deterministic components control eligibility, validation, authorization, persistence, and reporting. A framework is considered suitable only when its responsibilities are consistent with this boundary. For example, an LLM may propose a tool call or a code translation, but a schema validator, authorization rule, or target-side validator must determine whether the proposal can proceed. Components that merely forward data, provide infrastructure, or record observations are evaluated according to the contracts they expose rather than according to the quality of the final application output [4], [5], [34], [38]–[40]

### *B. Cross-Cutting Architectural Decisions*

Three decisions constrain the comparison across applications. They are treated as evaluation inputs rather than as framework-specific advantages.

***Shared LLM Gateway.*** Model access is centralized behind a shared internal gateway and client boundary where the application design requires common authentication, provider abstraction, timeout and retry policy, and usage capture. Individual components should not distribute provider credentials or implement incompatible model-access conventions. The gateway boundary is evaluated for response-shape consistency, requested and returned model identity, token accounting, failure propagation, and the possibility of provider-specific features such as streaming or function calling.

Centralization improves comparability because model identifiers, token usage, and latency can be recorded using a common representation. It also creates a shared dependency: gateway outage, quota exhaustion, or response-shape changes can affect several components simultaneously. The analysis therefore treats provider portability as a benefit only when it does not obscure the model configuration used for a result.

***Observability and Evaluation Conventions.*** Observability uses a common trace model based on OpenTelemetry conventions, with local extensions where an application-specific decision or retrieval event is not represented by the standard schema. Correlation identifiers are carried across component boundaries so that a final answer, action, or migration artifact can be related to the retrievals, model calls, validation stages, and failures that produced it. Telemetry is treated as diagnostic evidence and should not become a hidden acceptance condition unless the application explicitly requires it [35]–[37].

Evaluation records are structured by axis rather than reduced immediately to a composite score. A score should identify the measured axis, value, value type, evidence source, configuration or arm, sample size, denominator where applicable, run identifier, model identifier, and timestamp. Outcome metrics, compliance metrics, operational observations, human annotations, and judge-based scores are reported separately because they answer different questions.

The evidence hierarchy is conservative. Deterministic checks and execution outcomes provide stronger support for acceptance than human or model-generated judgments. Human labels can define a useful reference set but may reflect author or annotator assumptions. LLM-as-judge results can support triage and prioritization, but they are not treated as independent proof of correctness or as a deployment gate. Reported improvements require a meaningful baseline or control arm and sufficient repeated measurements to distinguish an effect from run-to-run variation [35]–[38].

***Synchronous and Durable Asynchronous Execution.*** Execution semantics are selected according to the interaction model. The chatbot exposes a synchronous turn boundary because a developer is waiting for a response. The automation system requires a durable asynchronous boundary because a workflow may remain paused for human approval for hours or days and must resume after worker replacement. The migration system processes a local run and records artifacts, with workflow-level recovery bounded by its configured repair and escalation policy.

The distinction affects framework fit. A synchronous conversational graph prioritizes low-latency state transitions and checkpoint persistence. A durable event-driven workflow prioritizes replay safety, external signals, idempotency, and recovery from worker failure. A batch-like migration workflow prioritizes reproducibility, dependency-aware ordering, bounded retries, and explicit terminal results. The same orchestration framework may appear in more than one application, but its suitability cannot be inferred without considering these different execution contracts.

### *C. Evaluation Criteria*

Each selected framework or implementation approach is evaluated against the criteria in Table I.

The criteria are applied qualitatively to framework fit and quantitatively where an evaluation experiment provides a valid measurement.

TABLE I. Criteria For Component-Level Framework Evaluation

| Criterion | Evaluation focus |
|---|---|
| Capability coverage | The extent to which the approach supports the component's required functions, including control flow, retrieval, model interaction, integration, validation, or state management. |
| Boundary coherence | The extent to which responsibilities remain explicit and non-overlapping, so that a component does not silently assume another component's decision authority. |
| Reliability and failure handling | Support for retries, timeouts, degraded modes, durable recovery, idempotency, bounded loops, and explicit terminal outcomes. An empty result, an unavailable dependency, and a successful result must remain distinguishable. |
| Security and governance | Support for authentication, authorization, approval before side effects, protection of untrusted content, secret handling, provenance, auditability, and controlled policy changes. |
| Integration complexity | The number and strength of contracts, services, adapters, version constraints, network boundaries, and process-level interactions introduced by the approach. |
| Observability and evaluation | The ability to record inputs, decisions, model and prompt metadata, latency, token usage, validation outcomes, and failure causes in a form that supports diagnosis and comparison. |
| Maintainability and portability | Clarity of ownership, testability, replacement cost, provider independence, and the extent to which application policy is separated from framework-specific behavior. |
| Operational efficiency | Model calls, token usage, latency, storage, deployment dependencies, and recovery overhead where these quantities are available. |

A high capability rating does not compensate for an unresolved security boundary, and a small efficiency improvement does not establish overall suitability when it reduces coverage or evidence quality. Recommendations are consequently conditional on the stated workload and risk model.

### D. *Evidence Sources and Assumptions*

The analysis distinguishes four kinds of evidence without treating any attached project artifact as a bibliographic reference. First, architectural evidence describes intended component responsibilities, considered alternatives, and cross-cutting constraints. Second, implementation evidence describes behavior present in the systems, including contracts, control flow, failure handling, and integration paths. Third, experimental evidence consists of controlled comparisons against a baseline or alternative under stated conditions. Fourth, evaluative evidence consists of tests, deterministic checks, execution outcomes, human labels, or other procedures used to assess a result.

Claims are reported at the level supported by the evidence. A tested invariant supports a claim about that invariant under the tested conditions; it does not establish production reliability. A structural similarity score supports a claim about similarity to the selected reference representation; it does not establish semantic equivalence. A measured latency or token count describes the evaluated configuration and workload; it does not establish a universal framework property. Where an alternative was only considered but not implemented or measured, it is described as a design option or hypothesis rather than as an observed result.

The evaluation assumes that the compared systems are assessed in their intended application contexts: the chatbot as an internal assistant over a constrained technical corpus, the automation system as an event-driven workflow with potentially long human-approval pauses, and the migration system as a local tool with partial Oracle-language coverage and target-side PostgreSQL validation. These assumptions exclude claims about open-domain conversational quality, arbitrary enterprise workflow scale, and complete Oracle-to-PostgreSQL behavioral equivalence.

The quantitative results have different experimental designs. Chatbot retrieval and routing results are reported with the evaluated corpus, dataset, repeats, metrics, and baseline. Migration results are reported with the sampled input pairs, decomposed tasks, strategy baselines, validation levels, and resource measurements.

For Automation, quantitative evidence is drawn from a frozen 40-case labelled triage set evaluated over repeated runs and a 29-case labelled runtime-guardrail set, with classification, routing, escalation, precision, recall, false-block rate, and reason-code accuracy reported separately according to their evidential role.

***Experimental Configuration.*** The chatbot retrieval comparison used an 80-node technical corpus, three repeated runs for the storage comparison, and a fixed retrieval configuration. The Automation evaluation used a frozen 40-case labelled triage set with five repeated runs for the principal classification comparisons and a separate 29-case labelled guardrail set. The migration evaluation used 1,000 input pairs decomposed into 1,448 tasks with seed 42; each principal strategy configuration was executed once, with up to 20 workers. These settings are reported to delimit the scope of the quantitative comparisons rather than to imply cross-application metric equivalence.

### E. *Component Contracts and Integration Analysis*

The component comparison is completed at the contract level before framework-level conclusions are drawn. For each application, the analysis identifies the data entering and leaving each component, the component that owns each decision, the failure states that can cross the boundary, and the evidence retained for later diagnosis. This prevents a framework from receiving credit for a capability that is actually supplied by application-owned policy or by an adjacent component.

The principal contract properties are type safety, explicit failure representation, provenance, authorization context, idempotency, and versionability. Typed results distinguish success from empty output and failure. Provenance identifies the source of retrieved context or the rule applied to a generated artifact. Authorization context binds a side effect to a principal, action, and approval state. Idempotency prevents retries or redelivered events from creating duplicate external effects. Versionability preserves the prompt, rule, model, analyzer, or configuration state needed to reproduce an outcome.

Integration analysis also considers negative paths. A component is not considered reliable merely because its success path is well defined. The comparison asks whether an unavailable store fails closed where required, whether invalid model output is rejected, whether missing dependencies lower the verification level, whether untrusted text is isolated, whether an approval can be reused incorrectly, and whether telemetry or an optional external service can fail without corrupting the authoritative result.

The resulting method separates five analytical layers: architecture description, component contract, observed behavior, quantitative evaluation, and interpretation. This separation supports the comparative sections that follow and limits conclusions to the claims supported by the available evidence. Fig. 1 illustrates the common architectural principle applied across the three evaluated systems: probabilistic components generate proposals, while deterministic components retain authority over validation, authorization, persistence, and final outcomes.

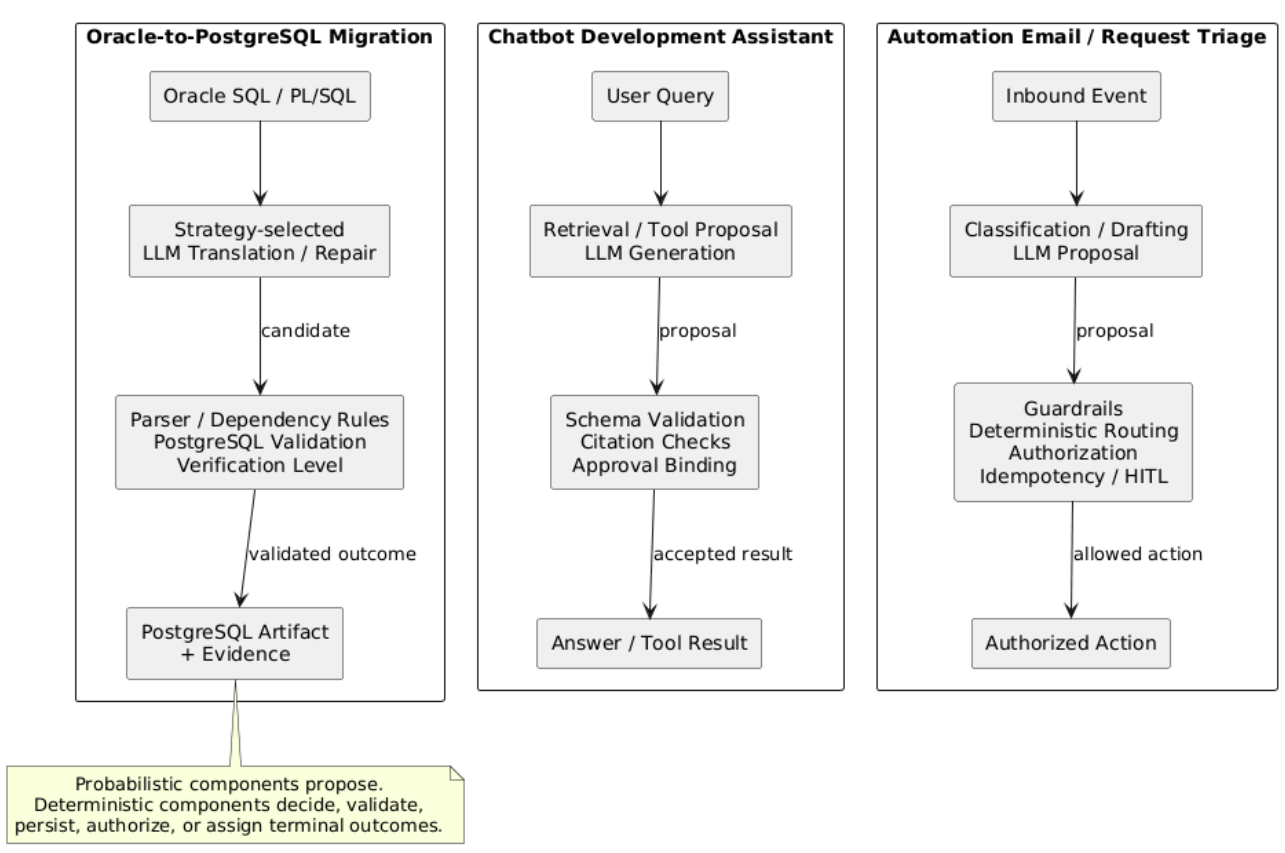

Fig. 1. Common probabilistic–deterministic control boundary across the chatbot, automation, and Oracle-to-PostgreSQL migration applications.

As shown in Fig. 1, the three applications differ in their probabilistic tasks, but they share the same control principle: generated proposals cross an explicit deterministic boundary before they can become accepted answers, authorized actions, or validated software artifacts. This common boundary provides the basis for the component-level comparisons in the following sections.

## III. Documentation-Grounded Chatbot Development Assistant

### A. Application Context and Requirements

The first application is an internal documentation-grounded development assistant intended for software engineers who need reliable answers about engineering standards, environment setup, APIs, and repository practices. The system objective is to support reduction of search and coordination effort for recurring technical questions while preserving source traceability and operational safety [17]–[25], [34].

The architecture documents define a five-component decomposition: conversation orchestration, retrieval, LLM and prompt management, tool integrations, and memory/evaluation/observability. In this decomposition, retrieval is modeled as a callable tool rather than as an independent branch, which allows a uniform request path for indexed documentation and live enterprise tools.

From a requirements perspective, the assistant serves three recurrent enterprise needs. First, it must provide grounded answers from an internal corpus with explicit citations that can be validated by users. Second, it must query live systems (for example Jira, repository search, and Confluence) when indexed data are insufficiently fresh or out of scope. Third, it must preserve operational controls for side-effecting actions through typed contracts and explicit approval gates. The principal functional requirements for this application are as follows.

- Context-grounded answering: the assistant shall generate responses from indexed internal documentation and return source-linked citations when evidence is available.
- Selective retrieval and tool use: the orchestrator shall decide whether to invoke the documentation retrieval tool, a live external tool, clarification, or direct response generation, instead of forcing retrieval for every turn.
- Uniform tool contract: all tool executions (including documentation retrieval) shall produce normalized typed results and typed failures, enabling deterministic downstream handling.
- Durable multi-turn state: conversation control state shall persist across turns and process restarts to support thread continuity and reproducibility.
- Human oversight for side effects: side-effecting tool calls shall be explicitly declared and gated via a human-approval path before execution.
- Observability and evaluation traceability: turns shall emit structured telemetry for retrieval, tool, and LLM events, and store comparable score records without collapsing them into a single opaque quality metric.
- Security boundary hardening: untrusted retrieved text shall be fenced/neutralized before prompt use, and policy-violating prompt behavior shall not bypass deterministic controls.

The non-functional requirements implied by the same evidence are: bounded integration complexity (single shared gateway and contract-first boundaries), reliability under partial failures (typed degraded modes instead of fabricated completions), maintainability via explicit component ownership, and auditability through structured traces and testable architectural invariants. Two assumptions delimit the interpretation of this section. First, this application is evaluated as an internal enterprise assistant with constrained corpus and controlled integrations, not as a general-purpose conversational agent. Second, some architecture artifacts remain marked as proposed, whereas repository tests document behavior already enforced in code; therefore, implemented constraints are treated as stronger evidence than unverified architectural intent.

## IV. Chatbot Architecture and Component Interactions

Fig. 2 presents the evaluated component architecture of the documentation-grounded development assistant and the interaction boundaries among conversation orchestration, retrieval, LLM and prompt management, enterprise tools, persistent state, and observability. The architecture separates probabilistic generation and routing from deterministic validation, approval, state persistence, and citation checking.

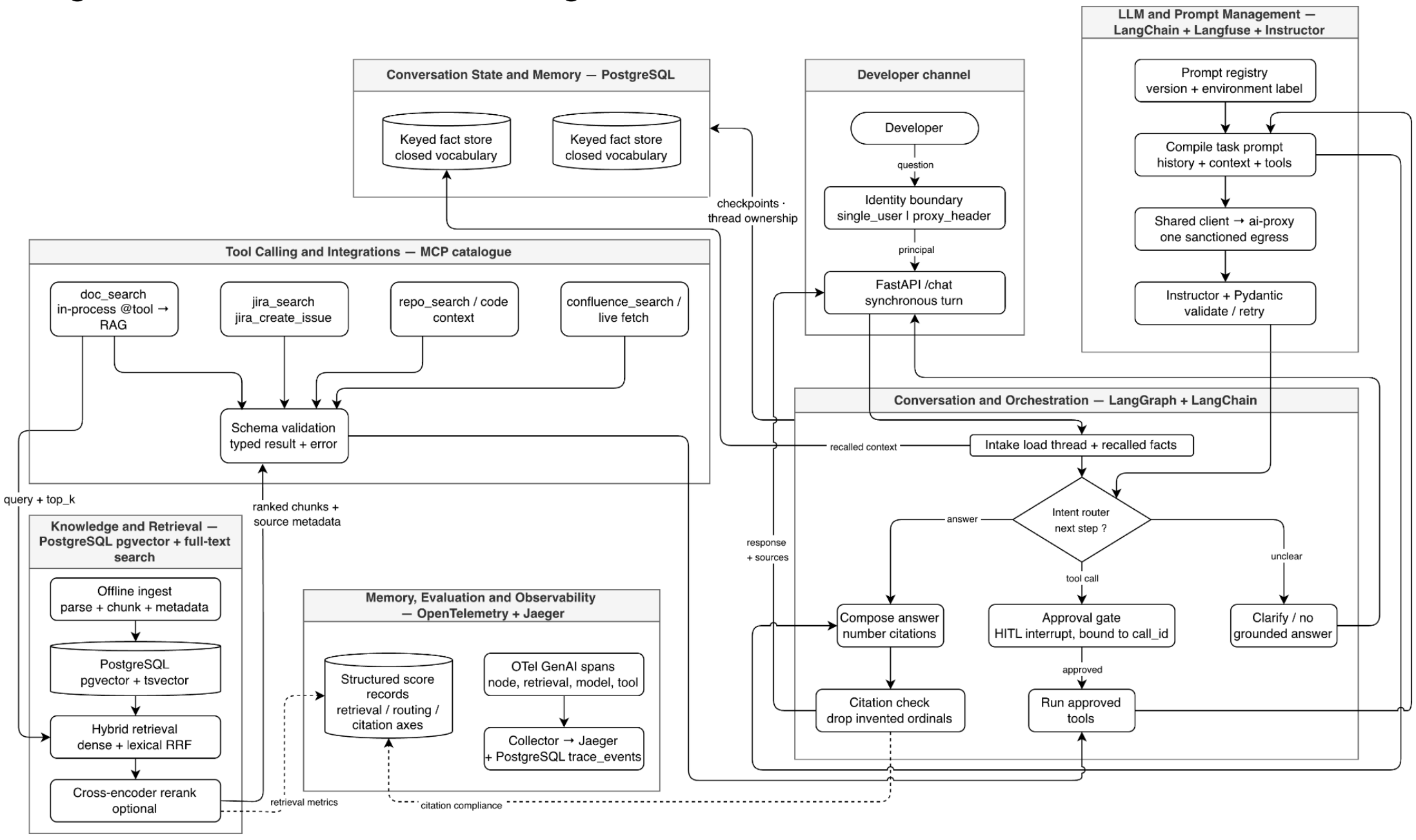

Fig. 2. Component architecture and interaction flow of the documentation-grounded development assistant.

As shown in Fig. 2, LangGraph provides the principal conversational control flow, while retrieval, model access, tool execution, persistent state, and evaluation remain separately owned components connected through explicit contracts. In particular, indexed-document retrieval is exposed through the same tool-oriented interaction boundary as live enterprise sources, while checkpoint persistence and citation validation remain deterministic controls outside the language model.

### A. Conversation and Orchestration

Conversation and orchestration define the control plane of the chatbot: they decide how each user turn is routed, how multi-step execution is coordinated across components, and how dialogue state is persisted across turns. The orchestration alternatives considered for the chatbot are summarized in Table II.

TABLE II. ORCHESTRATION FRAMEWORKS CONSIDERED FOR THE CHATBOT

| Framework | Role considered in planning | Strengths considered | Limitations | Status |
|---|---|---|---|---|
| LangGraph | Primary orchestrator | Explicit state graph, conditional edges, checkpointing, interruption support, recovery semantics for multi-turn workflows | Higher implementation complexity than linear chains | Selected as primary orchestration runtime |
| LangChain | Supporting orchestration library | Broad ecosystem for model adapters, message abstractions, and tool interfaces; useful as building blocks inside graph nodes | Linear chains alone are insufficient for durable branched control flow | Selected as supporting layer, not standalone orchestrator |
| OpenAI Agents SDK | Alternative orchestrator path | Fast setup, built-in session and guardrail primitives | Stronger provider coupling and weaker fit for complex branch control | Not selected; documented alternative |
| CrewAI | Multi-agent collaboration pattern | Useful for role-based agent collaboration tasks | Weaker fit for deterministic single-dialog routing with strict state control | Not selected |
| Temporal | Optional durability extension | Strong long-running workflow durability and retry semantics | Additional operational overhead for current short synchronous turns | Not selected; deferred unless long-running execution becomes required |

The final framework choice is therefore a layered combination: LangGraph for orchestration control flow, LangChain for reusable node-level abstractions, and PostgreSQL-backed checkpoint persistence for durable thread state. This combination is suitable for the evaluated requirements because it jointly satisfies: (i) explicit and inspectable routing control across retrieval, generation, clarification, and tool actions, (ii) restart-safe state continuity, and (iii) integration of human-approval interruptions without collapsing control logic into prompt behavior [18]–[23].

The orchestration scope is intentionally narrow. It owns turn intake, state loading, routing, loop control, and persistence; it does not own retrieval internals, prompt authoring, tool implementation logic, or observability backend operation. This boundary is important for component-level maintainability because it preserves single-responsibility contracts among the five chatbot components.

At the interface level, conversation orchestration consumes retrieved context, model outputs, tool results, and recalled memory, then emits structured requests to retrieval, generation, and tool-execution components while also emitting turn events for observability and evaluation. This architecture enables a uniform control path for both indexed documentation and live enterprise integrations, while keeping policy checks and failure handling at deterministic boundaries.

Documented failure modes for this layer include state-store unavailability, routing misclassification, peer timeouts, runaway loops, and checkpoint schema drift. Corresponding mitigation patterns are fail-closed behavior for persistence failures, explicit iteration limits, typed error propagation, and observability hooks for node-level diagnosis. As a result, the orchestration design prioritizes reliability and auditability over minimal implementation effort.

### B. Knowledge and Retrieval (RAG)

Knowledge and retrieval define the indexing and query pathways that ground the chatbot's answers in internal technical documentation. This component must efficiently surface both semantic matches for paraphrased questions and exact-identifier matches for configuration terms, while preserving metadata for source citation and freshness tracking. Tables III and IV compare the retrieval-framework and vector-store alternatives, respectively.

TABLE III. RETRIEVAL FRAMEWORK OPTIONS

| Framework | Strengths considered | Limitations | Status |
|---|---|---|---|
| LlamaIndex | Purpose-built retrieval abstractions and support for hybrid retrieval pipelines | The implemented query pipeline uses explicit SQL and PostgreSQL search primitives; retaining LlamaIndex as the retrieval runtime would add an unnecessary abstraction layer | Not selected; optional ingest dependency only |
| Haystack | Retrieval pipeline composition and support for multiple backends | Adds a second retrieval stack without a demonstrated capability requirement in this application | Not selected |
| LangChain | Integrates retrieval results with the surrounding orchestration and tool interfaces | General-purpose abstractions do not by themselves provide the required retrieval strategy or storage layer | Selected; used for the in-process tool boundary |

The final framework combination is therefore: retrieval as a callable tool in the orchestrator's catalogue, wrapping a hybrid dense-vector + sparse-lexical query pipeline fused with Reciprocal Rank Fusion (RRF), backed by PostgreSQL pgvector and native full-text search. This combination is suitable for the evaluated requirements because it satisfies three critical requirements: (i) selective invocation aligned with orchestration control flow, (ii) unified result envelope for both indexed and live sources, and (iii) dual-modality search (semantic + exact-match) appropriate for technical documentation containing both prose and configuration identifiers [17]–[25].

TABLE IV. VECTOR-STORE OPTIONS

| Vector store | Strengths considered | Limitations | Status |
|---|---|---|---|
| PostgreSQL pgvector extension | Co-located with the existing database; supports hybrid search with PostgreSQL full-text search; avoids an additional operational service for the current corpus | Not purpose-built for vector workloads; may show performance or feature limits at larger scale | Selected |
| Dedicated vector store (Qdrant, Weaviate, Pinecone) | Purpose-built approximate-nearest-neighbor indexing, richer filtering and hybrid-search APIs, and vector-specific operational features | Adds a second stateful service to operate | Not selected; migration path if pgvector shows a concrete performance or capability gap |

The retrieval scope is tightly scoped to the query pipeline and ingestion mechanics, not to usage decisions. It does not decide whether retrieval should run (that is orchestration's routing responsibility); it does not compose retrieved chunks into prose (that is LLM and prompt management's responsibility); and it does not measure retrieval quality or operate the evaluation backend (that is observability's responsibility).

At the implementation level, dense retrieval uses embeddings via the shared LLM gateway, lexical retrieval uses PostgreSQL full-text search with weighted section/body ranking, and Reciprocal Rank Fusion merges the two ranked lists at the position level to avoid score-calibration coupling. Metadata fields — source URL, document ID, section, and last-update timestamp — are mandatory on every chunk and travel through the typed result envelope to downstream components for citation reconstruction and freshness assessment.

Integration between retrieval and the tool-calling layer is via an in-process LangChain @tool wrapper for the current demonstration. The MCP server pathway remains a future deployment option. This design preserves the architectural principle that retrieval, like other tools, is invoked only when the orchestrator decides a query needs it.

Documented failure modes include vector-store unavailability, index staleness, embedding-gateway outage, and missing metadata. Mitigation patterns include graceful degradation to lexical-only search when dense retrieval fails, explicit flags for chunks with incomplete citations, and typed error codes propagated back to orchestration so failures do not collapse into empty-result silence. A production-scale corpus is not available for tuning thresholds on the current platform, so all measurements inherit this caveat.

### *C. LLM and Prompt Management*

LLM and prompt management provide the controlled generation boundary between conversation orchestration and the language model. This component is responsible for selecting versioned prompts, filling them with conversation and retrieval context, exposing the currently available tool schemas to the model, validating model outputs, and isolating provider-specific behavior from the rest of the application.

Tables V–VIII summarize the evaluated prompt-registry, prompt-compilation, structured-output, and provider-abstraction options.

TABLE V. PROMPT REGISTRY OPTIONS

| Framework | Strengths considered | Limitations | Status |
|---|---|---|---|
| Langfuse | Versioned templates, environment labels, runtime retrieval, and self-hosting; prompt versions can be correlated with evaluation data | Introduces an operational dependency | Selected |
| Portkey | Combines prompt management and gateway capabilities | Duplicates gateway responsibilities and creates competing ownership of prompts | Not selected |
| MLflow Prompt Registry | Mature experiment and artifact model | Less natural fit for continuously served prompts and operational tracing | Not selected |

TABLE VI. PROMPT COMPILATION AND MODEL ABSTRACTION OPTIONS

| Framework | Strengths considered | Limitations | Status |
|---|---|---|---|
| LangChain | Provides reusable prompt compilation and model-abstraction primitives without requiring it to own the agent loop | Must be constrained to compilation and model access; its automatic execution patterns would weaken orchestration control | Selected as a supporting layer |
| DSPy | Declarative signatures and prompt optimization against evaluation data | Optimization belongs to evaluation, while this component requires an explicit versioned prompt catalogue | Not selected |
| LlamaIndex or Haystack | Prompt templates and structured-output support | Adds coupling to an index or pipeline framework without a demonstrated need in this layer | Not selected |

TABLE VII. STRUCTURED-OUTPUT ENFORCEMENT OPTIONS

| Framework | Strengths considered | Limitations | Status |
|---|---|---|---|
| Instructor with Pydantic validation | Schema validation and bounded retry-with-feedback; works with an OpenAI-compatible endpoint | Adds a separate validation dependency | Selected |
| LangChain structured output | Already present in the surrounding stack | Combines compilation and validation, reducing separation of concerns and retry transparency | Not selected |
| Guardrails AI | Broader validation and corrective-generation capabilities | More execution and configuration machinery than schema conformance currently requires | Not selected |

The final design is a layered combination of Langfuse for prompt registration, LangChain for prompt compilation and model abstraction, Instructor with schema validation for output enforcement, and the shared internal client for model access through the enterprise gateway. These choices are suitable together because each framework has a bounded responsibility and the interfaces between responsibilities remain explicit. Prompt versioning is separated from prompt rendering; rendering is separated from provider access; and output validation is separated from tool execution and conversation control [12]–[16], [35]–[37].

The selected combination also fits the application's governance requirements. Langfuse provides the selected versioned prompt catalogue with environment labels and version identifiers. The compiler receives the task, conversation history, retrieved context, and current tool descriptions from orchestration, but it does not execute tools or own dialogue state. The model-access layer centralizes authentication, retry and timeout policy, model selection, and usage capture. Finally, schema validation converts an untrusted model response into a response that orchestration can process deterministically; if validation retries are exhausted, the request is returned as an explicit failure rather than an ambiguous completion.

TABLE VIII. GATEWAY AND PROVIDER-ABSTRACTION OPTIONS

| Option considered | Strengths considered | Limitation | Status |
|---|---|---|---|
| Shared internal client through the sanctioned enterprise gateway | Single egress, centralized authentication, retry and timeout policy, provider neutrality, and consistent usage capture | Depends on the availability and governance of the internal gateway | Selected |
| LiteLLM | Multi-provider normalization and fallback routing | Adds a second gateway layer and network hop where the sanctioned gateway already provides provider abstraction | Not selected |
| Portkey gateway | Gateway, prompt vault, and routing features | Duplicates selected prompt-registry and gateway responsibilities and increases vendor coupling | Not selected |
| Direct provider SDKs | Maximum direct control over provider-specific features | Spreads credentials, retries, and provider coupling across call sites | Not selected |

The component boundary is deliberately narrower than general-purpose agent frameworks. LLM and prompt management does not persist conversation history, perform document retrieval, execute external tools, manage cross-session memory, or operate the evaluation backend. This prevents prompt compilation from silently becoming a second orchestration layer and preserves the separation of responsibilities established by the chatbot architecture.

The main interaction sequence is: orchestration supplies a task and contextual inputs; the prompt registry supplies the selected version; the compiler renders the model input; the shared gateway performs the model request; and the output validator returns a structured result or a typed failure. Each call also produces telemetry containing prompt identity and version, model information, rendered input, response data, token usage, and latency. Telemetry is emitted independently of the user request path so an observability outage does not block answer generation.

The principal failure modes are unavailable prompt-registry data, gateway or provider failure, invalid model output, and model responses that request unavailable tools. The design addresses these risks through explicit prompt availability checks, centralized retry and timeout policy, bounded schema retries, dynamically supplied tool descriptions, and typed error propagation to orchestration. The architecture therefore favors provider portability, prompt auditability, and predictable integration behavior over the smallest possible number of dependencies.

### *D. Tool Calling and Integrations*

Tool calling and integrations form the boundary between the assistant and live enterprise systems. The component enables the assistant to obtain current information from issue tracking, source repositories, and knowledge platforms that may not be represented in the indexed documentation corpus. It also converts model-generated action proposals into validated requests, so the model does not directly control external systems.

The tool-integration alternatives and the selected deterministic validation boundary are summarized in Table IX.

TABLE IX. TOOL-INTEGRATION FRAMEWORKS AND VALIDATION APPROACHES

| Framework or approach | Role considered in planning | Strengths considered | Limitations | Status |
|---|---|---|---|---|
| Model Context Protocol (MCP) | Standard integration boundary for external tools | Language- and framework-independent; separates tool ownership from orchestration; supports discovery and independent tool servers | Requires server lifecycle, authentication, and operational management | Selected as the integration standard |
| LangChain tools | In-process doc_search wrapper and function interface | Simple integration with the orchestration stack; low overhead; convenient typed interface | The external MCP adapter package is incompatible with the selected LangChain Core version; external tools use the direct MCP SDK | Selected; limited to the in-process retrieval boundary |
| Provider-native function calling | Direct model-to-provider tool invocation | Minimal infrastructure and broad provider support | Creates provider coupling and places tool definitions in provider-specific request formats | Not selected |
| Trust model-generated calls without validation | Minimal execution path | No additional validation layer | Malformed, unknown, or unsafe calls can reach enterprise systems | Not selected |
| Boundary schema validation | Deterministic validation before execution | Rejects invalid arguments before external effects; produces typed failures | Schemas must be maintained as tools evolve | Selected |

The implemented integration model uses MCP as the provider-neutral boundary for live enterprise sources, with the direct MCP SDK responsible for discovery, invocation, reconnection, and shutdown. An in-process LangChain @tool wrapper exposes only the indexed-document search path. The LangChain MCP adapter package is not used because the evaluated releases either lost structured tool payloads or conflicted with the selected LangChain Core version. This preserves independently managed live tool servers without misrepresenting the current transport implementation.

The component exposes a reviewed catalogue of available tools and their input and output schemas. The catalogue also declares whether an operation is read-only or side-effecting. The orchestrator decides whether a tool should be called, but the integration component validates the selected name and arguments before execution, normalizes the result, and returns a typed success or failure. This implements the architectural principle that the language model proposes an action, whereas deterministic boundary logic decides whether that action is well-formed and eligible to run.

For the development assistant, live integrations address information that changes after documentation ingestion. Issue-tracking search supports current work-item status, repository search supports current source and change information, and live knowledge-platform retrieval supports freshness-sensitive pages. The indexed documentation pathway and live retrieval pathway can overlap, particularly for knowledge-platform content; the orchestration layer must select the appropriate source according to freshness and task requirements.

The selected boundary also supports human oversight. Read-only operations may be executed automatically when authorized by orchestration. Side-effecting operations must be marked as such and routed through a human-approval step before execution. Approval is associated with the specific proposed operation and its arguments, preventing a later or different request from reusing an earlier approval. In the current demonstration, the write-capable path is exercised in a controlled fixture mode; production side effects are not assumed to be enabled.

Results are normalized before they reach answer generation. Successful calls contain bounded structured data, while failures contain an explicit reason such as invalid arguments, unavailable upstream service, timeout, or unknown operation. This distinction is important for grounded answering: an empty result is not equivalent to an outage, and an outage must not be represented as evidence that can support a confident answer. Result truncation and schema normalization also limit accidental disclosure of excessive or sensitive upstream data.

The main failure modes are unavailable or slow external systems, malformed or nonexistent operations, missing approval for a side-effecting action, and oversized or sensitive result payloads. The corresponding controls are timeouts and typed failures, boundary validation, mandatory side-effect declarations, human approval, result normalization, and structured execution telemetry. Evaluation focuses on engineering invariants such as rejection before execution, completeness of read/write declarations, bounded result shape, and preservation of failure information; tool-selection quality itself belongs to orchestration evaluation.

### E. Memory, Evaluation, and Observability

Memory, evaluation, and observability provide the accountability layer for the chatbot. Memory supplies limited cross-session personalization, observability records how a turn was produced, and evaluation determines which properties can be checked reliably. These responsibilities are related but are not interchangeable: memory supports the user experience, traces support diagnosis and governance, and evaluation supports evidence-based comparison.

Tables X–XII summarize the memory, observability, and evaluation alternatives considered for the chatbot.

TABLE X. CROSS-SESSION MEMORY OPTIONS

| Framework or approach | Strengths considered | Limitations | Status |
|---|---|---|---|
| Keyed fact store in the application database | Small operational footprint; sufficient for a limited number of explicit user facts; easy to inspect and control | No automatic extraction, temporal reasoning, or conflict resolution | Selected |
| Mem0 | Purpose-built memory API with automatic fact extraction and updates | Adds a service and dependency for a requirement currently satisfied by explicit facts | Not selected; reconsider if personalization expands |
| Zep | Temporal memory and knowledge-graph capabilities | Greater operational and conceptual complexity than the current scope requires | Not selected; justified only for richer temporal personalization |
| Custom summarization pipeline | Flexible free-form session summaries | Adds another model call and introduces additional cost and quality risks | Not selected |

TABLE XI. OBSERVABILITY AND EVALUATION-PLATFORM OPTIONS

| Framework or approach | Strengths considered | Limitations | Status |
|---|---|---|---|
| Langfuse | Versioned prompt registry, datasets, and score-management capabilities | Tracing through the installed Langfuse v2 endpoint is not the implemented path | Selected |
| Phoenix | OpenTelemetry-native tracing and strong trace inspection | Its evaluation approach may encourage judge-primary assessment, which conflicts with the project's evidence hierarchy | Not selected |
| MLflow | Mature experiment tracking and evaluation support | Less natural fit for continuous conversational traces and online operations | Not selected |
| OpenTelemetry with a project-managed store | Portable trace standard and control over storage and schema | Requires the team to provide storage, querying, and visualization | Selected |

TABLE XII. EVALUATION-METRICS OPTIONS

| Framework or approach | Strengths considered | Limitations | Status |
|---|---|---|---|
| RAGAS and DeepEval | Reusable retrieval and answer-evaluation metrics | Optional evaluation dependencies; metrics do not replace labelled data or establish an automatic truth source for free-form answers | Selected; available through the optional evaluation extra |
| LLM-as-judge | Can screen qualitative responses and prioritize human review | Subject to judge bias and calibration drift; unsuitable as an acceptance gate | Selected; advisory triage policy only |

The final framework choice is deliberately lightweight. Cross-session memory is implemented as a keyed fact store rather than through Mem0, Zep, or an automatic summarization framework. This is suitable because the assistant's current personalization requirement is limited to a small set of explicit facts, such as stable user or project preferences. A thin key-value model avoids introducing a new reasoning layer, while its limitations remain visible: updates use a simple consistency policy and may not support same-session conflict resolution.

For observability, the architecture requires a common OpenTelemetry-based trace model across orchestration, retrieval, model calls, and tool execution. The implementation uses structured spans, a durable local database sink, and OTLP export to Jaeger for trace inspection. Langfuse is a verified prompt-registry dependency, but its v2 deployment is not the implemented trace-ingestion backend; prompt serving and tracing must therefore be reported as separate capabilities.

Evaluation is organized around separately reported axes rather than a composite quality score. Labelled datasets and control arms support verifiable outcomes such as retrieval recall, retrieval precision, and tool-call correctness. Static checks cover citation compliance, sensitive-data handling, and other deterministic properties. Latency, token usage, and cost are monitored as operational observations. Free-form helpfulness and tone do not have a reliable automatic oracle in this application, so human annotation may inform improvement but cannot serve as a definitive acceptance gate. An LLM judge may triage cases for review, but its output is advisory and must not gate deployment [25]–[27], [35]–[38].

The component receives turn events, retrieved-document identifiers, model and prompt metadata, and tool outcomes from the other chatbot components. It returns recalled facts to orchestration before generation and produces traces, score records, evaluation reports, and review queues for operators and developers. Prompt content, model identifiers, token counts, latency, retrieval evidence, and tool outcomes must remain correlated at the turn level so that a regression can be attributed to a component or configuration rather than inferred from a final answer alone.

The principal failure modes are unavailable telemetry storage, stale or incorrect memory, unmeasured evaluation thresholds, and overinterpretation of judge scores. The corresponding controls are non-blocking telemetry emission, explicit memory scope, repeat-derived per-axis variation estimates, control-arm comparison, and clear separation between verifiable outcomes, human annotations, and judge-based triage. This design makes uncertainty visible instead of converting missing evidence into a zero score or an unsupported quality claim.

## V. Chatbot Component Fit and Trade-Offs

The selected design uses broad framework capabilities where they support the control plane and model interaction, and specialized components where the application has stronger requirements for retrieval quality, tool isolation, memory scope, or evaluation accountability. The resulting fit is assessed against five criteria: capability coverage, boundary coherence, reliability, operational complexity, and evidence of successful implementation. Table XIII summarizes the resulting component-level fit and trade-offs.

TABLE XIII. Chatbot Component Fit and Framework Trade-Offs

| Component | Selected framework or approach | Capability fit | Main benefits | Main trade-offs and residual risks | Evidence status |
|---|---|---|---|---|---|
| Conversation and orchestration | LangGraph with LangChain support and PostgreSQL checkpoints | High fit for conditional routing, iterative execution, durable state, and human interruption | Explicit control flow; restart-safe thread state; clear integration points; model portability | More state-schema and operational complexity than a linear chain; database availability is critical; loop and migration controls are required | Architecture selected; core graph, persistence, and approval behavior tested |
| Knowledge and retrieval | Hybrid dense and lexical retrieval with rank fusion, co-located in PostgreSQL; in-process integration for the demo | High fit for technical documentation containing both paraphrases and exact identifiers | Combines semantic and exact-match retrieval; preserves citation metadata; avoids a second vector service at current scale | Two retrieval arms must be maintained; shared database resources may contend with checkpoints; larger-scale performance is not established; reranking is disabled by default after negative current-corpus results | Retrieval, ingestion, metadata, and ranking behavior tested; production-scale validation remains open |
| LLM and prompt management | Versioned prompt registry, LangChain compilation/model abstraction, schema validation, and shared enterprise model client | High fit because prompt governance, provider abstraction, and output validation are separate concerns | Prompt rollback and environment control; centralized access policy; structured outputs; provider portability | Prompt registry and enterprise gateway become critical dependencies; validation retries add latency; framework boundaries must prevent accidental tool execution | Core contracts and model/gateway behavior verified; operational deployment details remain environment-dependent |
| Tool calling and integrations | MCP direct-SDK transport, in-process LangChain doc_search wrapper, and deterministic schema validation | High fit for independently owned, extensible enterprise integrations | Language-neutral source boundary; reviewed catalogue; typed results and failures; explicit read/write classification | Server lifecycle, authentication, and permissions add operational work; the LangChain MCP adapter is unavailable for the selected dependency versions; per-user authorization remains an open concern | Tool discovery, validation, failures, and approval path tested; live credentials are not universally available |
| Memory, evaluation, and observability | Keyed fact store; OpenTelemetry traces with project-managed durable storage and Jaeger export; optional labelled metric evaluation | High fit for limited personalization and evidence-based system monitoring | Small memory surface; cross-component trace correlation; separate outcome and compliance measures; judge output remains advisory | Keyed memory lacks temporal reasoning; trace operations require ownership; free-form answer quality lacks a verifiable oracle; RAGAS and DeepEval require the optional evaluation extra | Memory, spans, score records, and offline evaluation paths implemented; human-quality evidence remains limited |

The clearest architectural fit occurs at the boundaries where the requirement is concrete and mechanically checkable. PostgreSQL-backed checkpoints directly address durable conversation state; metadata-preserving retrieval directly supports citations; schema validation directly constrains tool execution; and structured traces directly support diagnosis and evaluation. These choices are more defensible than selecting one framework for all responsibilities because each selected mechanism corresponds to a distinct failure mode or contract requirement.

The principal trade-off is complexity. Five components and several specialized frameworks create more contracts, deployment dependencies, and versioning decisions than a single agent loop. That cost is justified only because the application requires behavior that a simple loop would leave implicit: durable state, selective retrieval, live-system access, approval before side effects, source traceability, and measurable failure handling. The architecture reduces this complexity through explicit ownership, normalized result shapes, a shared model-access boundary, and common telemetry conventions.

The second trade-off concerns operational maturity. The implementation demonstrates the control flow, contract checks, retrieval behavior, memory store, and evaluation machinery, but some proposed infrastructure paths are not equivalent to production evidence. In particular, current measurements use a small documentation corpus and authored evaluation data, while live external-system access and sustained concurrent behavior remain environment-dependent or unmeasured. Consequently, the framework selection supports a strong prototype and evaluation baseline, but claims about production scalability, long-term memory quality, and end-user answer helpfulness require additional evidence.

Overall, the chatbot exhibits high component fit for its documented prototype objectives. LangGraph and LangChain provide the general control and model-interaction foundation; retrieval, MCP, the keyed fact store, and OpenTelemetry supply narrower capabilities that address enterprise-specific risks. The design is therefore suitable when auditability and controlled integration are more important than minimal dependency count, provided that unresolved authorization, scale, data-residency, and operational-ownership questions are addressed before production deployment.

## VI. Chatbot Reliability, Security, and Operational Risks

The chatbot's reliability and security model is based on deterministic boundaries around inherently probabilistic generation. The language model may propose an answer or tool call, but it does not decide whether an external operation is valid, whether a side effect may run, or whether an unavailable source can be represented as evidence. These decisions remain in typed contracts, orchestration, and tool-execution controls [1], [4], [15], [34]–[37].

Reliability depends first on preserving the distinction between a result, an empty result, and a failure. Retrieval and tool integrations return typed outcomes with explicit reason codes for conditions such as timeouts, unavailable indexes, unavailable upstream services, and invalid arguments. This prevents an outage from being transformed into an apparently grounded answer. State persistence is similarly fail-closed: when the PostgreSQL checkpoint store is unavailable, the system rejects or fails the turn rather than continuing without durable conversation state. Loop limits and structured error propagation bound failures within the orchestration path.

Security controls apply at the boundary between untrusted content, model input, and enterprise systems. Retrieved documents and tool payloads are treated as data rather than instructions: prompt-injection markers are neutralized and untrusted material is fenced before it is inserted into a model prompt. Tool results pass through a central normalization and redaction path before they can reach the model, traces, or checkpoints. This path removes values matching secret-related patterns and sanitizes provenance fields such as URLs and identifiers, reducing the likelihood that credentials or embedded access tokens are reproduced in a response or retained in operational records.

The most consequential control concerns side effects. Every tool declares whether it requires approval, and the orchestration layer suspends execution when a proposed side-effecting call is encountered. An approval is bound to the specific pending call and the initiating principal; stale, mismatched, unsolicited, or cross-user approvals are rejected. Read-only calls may proceed only through the same validated catalogue and schema boundary. Consequently, model-generated arguments are proposals subject to validation, rather than instructions that can directly invoke an enterprise system.

The principal chatbot reliability and security risks are summarized in Table XIV.

TABLE XIV. Chatbot Reliability, Security, and Operational Risks

| Risk | Consequence | Implemented control | Remaining limitation |
|---|---|---|---|
| PostgreSQL checkpoint-store outage | Loss of durable thread state or inconsistent approval handling | Fail-closed turn processing, typed errors, and persisted checkpoint contracts | Database availability remains a critical dependency |
| Retrieval or live-tool outage | An answer may appear to be supported by unavailable evidence | Typed failure results, timeouts, degraded retrieval handling, and prohibition of confident generation over missing results | Upstream availability and recovery time remain outside the assistant's control |
| Prompt injection in retrieved or live content | Untrusted text may attempt to alter model behavior | Neutralization and fencing of untrusted content before prompt assembly | Prompt isolation reduces risk but cannot prove that a model will never follow malicious text |
| Secret or sensitive-data exposure in tool results | Credentials or excessive upstream data may enter prompts, traces, or checkpoints | Central result normalization, redaction, truncation, and URL/identifier sanitization | Pattern-based redaction may not detect every sensitive value; source-system access policy remains essential |
| Hallucinated tool name or malformed arguments | Invalid requests reach enterprise integrations | Reviewed catalogue, typed input schemas, and validation before execution | Tool schemas and catalogue entries must be maintained as integrations evolve |
| Unauthorized or unintended side effect | External state changes without accountable approval | Mandatory read/write declaration; approval bound to the pending call and initiating principal | Per-user authorization for permission-scoped live data remains an unresolved deployment concern |
| Excessive iteration or repeated dependency failure | Latency and resource consumption increase without a useful answer | Explicit orchestration iteration limits, timeouts, and typed propagation | Thresholds require production-like workload evidence |
| Telemetry or evaluation-store outage | Reduced traceability and delayed diagnosis | Non-blocking telemetry emission and durable local trace/score storage | Observability degradation can conceal emerging quality or security regressions |
| Incomplete citations or stale corpus content | Users may over-trust incomplete or outdated documentation | Mandatory citation metadata, explicit completeness flags, and freshness fields | Corpus refresh policy and production-scale staleness behavior require further validation |

The implementation includes tests for approval binding, fail-closed grounding, prompt-injection handling, secret redaction, URL sanitization, typed failures, and tool validation. These checks provide evidence that the stated boundary controls are enforced in the current prototype. They do not establish equivalent protection under all providers, all enterprise identity models, or sustained concurrent production traffic.

The principal residual risks are operational and governance-related. Live integrations require a per-user authorization model before permission-scoped data can be exposed safely. The corpus requires an ownership and refresh policy to manage stale content. Retention, data residency, and access control for prompts, checkpoints, traces, and evaluation data must be defined before production deployment. Finally, the small authored datasets and absence of long-running production traffic limit claims about routing robustness, retrieval quality at scale, and the practical effectiveness of the human-approval process.

# VII. Automation Architecture and Component Interactions

## A. Workflow / Pipeline Orchestration

***Purpose.*** This component sequences an inbound request through the pipeline (read → classify → decide → act) as an explicit, inspectable flow rather than implicit call nesting; provides durable checkpointing so a crash resumes from the last completed step, not from the top; owns per-step retry and backoff policy; invokes the decision, validation, and authorization layers and branches on their verdicts; and owns the control-flow mechanics of a human-in-the-loop pause and resume — the pause itself, not the approval surface or the rules behind approval. Without this component, the other four capabilities exist in isolation with nothing to carry one request through them in order, retry the step that failed, or resume a run that was waiting on a human when a worker restarts.

***Internal Structure.*** The component is organized as a declared topology of steps and branch predicates: a classification/decision step, a draft-validation step, an authorization step, and a human-review step, connected by conditional edges (proceed / route-to-human / block). Two structurally separate execution concerns sit inside this one component: a control-flow layer that expresses the graph of steps and branches as an explicit, inspectable structure, and a durability layer that checkpoints progress at step boundaries so a crash or a multi-day human-approval pause resumes rather than restarts. Every outbound action carries an idempotency key so a resumed or retried run cannot produce a duplicate side effect.

***Interactions.*** Consumes: a normalized inbound event from the integrations component; a typed decision result (classification, routing, draft, escalation signal) from the agent decision layer; a guardrails verdict (pass/block/revise) from the monitoring component; an authorization verdict (allow/block/needs-approval) plus a durable state handle from the business-logic component. Produces: an enriched task context (raw message, channel, routing snapshot, revision context) to the agent decision layer; a draft-validation request to the monitoring component; a proposed action to the business-logic component, followed by a completion report once the integrations component has executed (or failed to execute) the authorized action; an authorized action command to the integrations component; and pipeline step events (span content) to the monitoring component.

***Framework Analysis.*** The central framework question for this component is how to combine an explicit, branch-and-loop control-flow representation with durable, resumable execution — because a run may legitimately pause for hours or days on human approval and must survive a worker restart while paused [47]. Table XV compares the principal orchestration and durable-execution alternatives considered for this requirement.

TABLE XV. Workflow Orchestration and Durable-Execution Options

| Option | Capability fit | Operational impact | Reliability | Risks & lock-in | When justified |
|---|---|---|---|---|---|
| LangGraph (control flow) + Temporal (durable execution) | Graph expresses decide/act/route-to-human/block as explicit edges; Temporal gives crash-survivable, resumable execution with built-in retry/backoff and event-sourced history | Two services to run and operate (a graph runtime plus a durable-execution cluster/worker) | Strong: durable per-activity retry, replay-safe history, effectively-once side-effect behavior under retries when the target action and idempotency mechanism satisfy the stated assumptions | Two frameworks with independent release cycles; the two do not compose literally — see measured finding below | Default whenever a run must survive a multi-day human-approval pause and worker restarts, and step-level retry/backoff must be durable, not just in-process |
| LangGraph alone, with its own checkpointer, no separate durable-execution engine | Graph structure identical; checkpointer persists graph state and covers crash-resume for short flows | One fewer service to operate | Weaker: checkpointer-based resume is not the same guarantee as a durable workflow engine's event-sourced history for long-lived, multi-day pauses and effectively-once side-effect behavior under appropriate idempotency assumptions | Lighter, but the team hand-builds durable replay and retry guarantees that otherwise require additional application-level handling | When runs are short, side effects are naturally idempotent at the target system, and no step pauses for a human-timescale duration — a reasonable lighter-weight choice for a prototype or a low-stakes internal tool |
| Prefect / Airflow (data-pipeline scheduler) | Mature scheduling, retries, and observability for batch DAGs | Familiar to data-engineering teams already running one of these schedulers | Good for scheduled, batch-shaped work; awkward for a flow that pauses mid-run on an unscheduled human verdict and resumes on an external signal | Adds a second orchestration paradigm (scheduled-batch) on top of an event-driven, per-request flow | Only if the automation is a scheduled bulk job (e.g. nightly batch reclassification) rather than a per-request, human-gated flow triggered by inbound events |
| An open-ended agent loop as the orchestrator (the agent "figures out" the steps) | Minimal to stand up | None — no separate service | Removes the deterministic gating (authorization verdict, HITL pause) this component exists to enforce; introduces a second thing that "orchestrates" beside the graph | Lets the model improvise control flow around an authorization gate, defeating the gate's purpose | Never as the pipeline spine when a deterministic authorization gate must be enforced before any side effect; acceptable only for a non-actioning, read-only exploration agent |

Measured finding: In the evaluated LangGraph–Temporal version pairing, the tested LangGraph graph could not execute successfully inside Temporal's deterministic workflow sandbox. A three-node LangGraph graph, run inside a sandboxed Temporal workflow, fails with Cannot access os.environ.get from inside a workflow because LangGraph reads environment state while building the graph; run against an unsandboxed workflow runner, the workflow task still fails during activation. Because Temporal checkpoints at activity boundaries, whatever sequences the steps has to be the thing Temporal is executing — wrapping the entire LangGraph graph inside a single Temporal activity would make it durable as one unit, but would forfeit "resume from the last completed step," the exact property the durable-execution engine was chosen to provide. The working resolution is to declare the pipeline's topology and every branch predicate once, in a pure, framework-independent form; let LangGraph genuinely execute the non-durable path (so the graph stays inspectable and can be rendered); let Temporal execute the durable path against the same predicates; and add a consistency test that both executors produce identical results on identical input, so the two paths cannot silently diverge. This is a structural composition finding about combining a graph-orchestration library with a durable-execution engine specifically, not an argument against either framework alone — teams adopting this combination should plan for this compensating layer rather than assuming the two frameworks nest directly. The general methodological point is broader than this one pairing: both frameworks were independently well-justified on their individual merits, and the incompatibility was invisible to that individual-merits analysis — it existed only in the combination, and a two-framework composition claim of this kind is only testable by building it, not by evaluating each framework's documentation separately.

Measured finding: a paused run surviving a full worker replacement. Against a real Temporal deployment, a workflow paused awaiting human approval was shown to survive the worker process being replaced entirely: a new worker reconstructed the run from durable event history, answered a status query for it, and completed it on approval with the agent decision layer consulted exactly once rather than twice on resume. This is the concrete evidence behind the "durable execution engine" row's reliability claim above, not an assumption.

***Assumptions / Open Questions.*** None of the framework configurations evaluated in this study provided an integrated human-approval notification surface as part of the tested control-flow path. LangGraph's interrupt() primitive and Temporal's signal/update mechanism both provide the pause and resume control flow, but neither provides a default UI or notification channel (email, dashboard, chat) for the human to act on. Question to confirm: for a given deployment, which existing enterprise surface (ticketing system, chat tool, dedicated review UI) should receive the pause notification, and does that choice change the recommendation above? The evaluated configurations therefore exhibit a common notification-surface gap, while the appropriate enterprise surface for closing that gap remains deployment-specific.

The LangGraph/Temporal composition finding above was measured against one version pairing of each framework. Whether a future release of either framework changes the sandbox-incompatibility behavior is not established; teams adopting this combination should re-verify the finding against the specific versions in use rather than assuming it is permanent. Checkpointer-only durability (LangGraph alone) was not independently benchmarked against Temporal for the multi-day-pause case — the recommendation above is reasoned from each framework's documented guarantees, not from a side-by-side measurement of resume latency or failure-mode coverage.

*B. Agent Decision Layer*

***Purpose.*** The agent decision layer is the pipeline's only probabilistic step: it turns an unstructured inbound message into a structured proposal — a classification, a routing target, a draft response, and an escalation signal — that downstream deterministic layers can consume without themselves having to interpret natural language. Its output is a proposal, never a decision; whether the proposal is authorized and executed is delegated entirely to the components that follow it [18]–[23], [28]–[30], [36]–[40].

***Internal Structure.*** The layer is typically decomposed into: a classification step producing a typed label with a confidence signal; a routing-resolution step mapping the label to a target queue or department; a drafting step producing a department-scoped response; and an escalation-signaling step that distinguishes "low confidence but a complete proposal exists" (routing and drafting still run so a human reviews something concrete) from "no usable proposal" (no routing match, drafting is not structurally possible).

Whether routing resolution is itself an LLM step or a deterministic lookup is a first-order design choice, not an implementation detail — it determines whether routing accuracy can be evaluated and audited independently of classification accuracy.

***Interactions.*** Consumes: an enriched task context from the orchestration component (raw message, channel, a versioned routing-rules snapshot, and a revision flag/context if the run is on a second pass after a guardrail or human rejection).

Produces: a typed decision result (classification, routing target, draft, escalation flag/reason) returned to the orchestration component, which forwards it to the business-logic/authorization component and, on an escalation or needs-approval verdict, to the human-in-the-loop pause point owned by the orchestration layer.

***Framework Analysis.*** Measured finding: framework-side hand-off, not model-side — measured in a standalone run of the framework, not in the shipped decision layer. The deciding argument for choosing a role-based multi-agent framework over a function-calling-based agent SDK was that the former's sequential hand-off is framework-side Python, not a model-side function call — a claim that was measured directly rather than taken from vendor documentation, in a dedicated exploratory run built specifically to exercise the framework's hand-off. A two-agent sequential CrewAI run put exactly two chat-completion requests on the wire, neither carrying tools, tool_choice, or functions, confirming that, in the evaluated CrewAI configuration, the sequential hand-off was executed in framework-side Python rather than through a model-side function call. This demonstrates that such a property can be verified experimentally for a specific role-based framework and deployment path rather than inferred from framework documentation alone.

Measured finding: a shipped component can adopt an agent framework's prompt convention while never executing the framework itself, and still pay its full dependency cost. In the reference build, the decision layer's classification and drafting steps call the LLM client directly from plain functions; the framework's own hand-off/execution entry point is constructed (its role objects are built) but never invoked, confirmed by asserting the framework's package never enters the running interpreter's loaded-module set across a live decision run.

Every live model call in the campaign measured for this report went through that direct path, not through the framework. What did reach the reference implementation is the framework's prompt convention — its role/goal/backstory framing — hand-inlined as literal prompt text in the classification and drafting prompts, authored independently of the framework's own objects. The framework was nonetheless fully present as a dependency, capping four shared package versions for other components (see the dependency-pinning finding below) for a code path that never runs.

The general lesson for framework-fit analysis: verifying that a framework is imported, or even that its objects are constructed, is not evidence that it is executed — only tracing whether its package enters the running process during a live call settles that, and the answer can diverge from what the architecture document describes even when the document and the import statement agree. Measured finding: the classification-boundary defect was deployment-dependent in the evaluated experiments rather than attributable solely to the framework or prompt formulation. Two labels in a classification vocabulary (a general-catch-all intent and a "no confident match" sentinel) competed on ambiguous input on a small, cheap model deployment: one hand-written test case (a clearly out-of-domain message) was misclassified as the general catch-all in every repeat. Table XVI compares the agent-decision-layer alternatives and the conditions under which each design is justified.

Two rounds of intent-description prompt tuning failed to resolve it — one rewrite also destabilized a previously stable, unrelated case — and a 40-case, 5-repeat labelled measurement (see Section VII-B) confirmed the effect was real, not a smoke-test artifact, and additionally surfaced a second, unambiguous misclassification the four hand-written cases had missed. Re-running the identical, unedited configuration on a larger model deployment resolved the originally targeted case correctly in 5/5 repeats, while the same rewrite that helped the smallest deployment measured as a regression on the mid-tier deployment. The operative finding is therefore about evaluation methodology for any agent decision layer, independent of which framework hosts it: a classification-quality defect measured on one model deployment should initially be treated as deployment-conditioned evidence rather than as a general property of the prompt or intent vocabulary, until a second, materially different deployment is tried. Treating a single-deployment measurement as a property of the pipeline design is a documented, reproducible mistake — and, per Section VII-B, its cost was six rounds of prompt/description rewriting before a materially cheaper experiment (swapping the model, holding everything else fixed) settled the question.

Measured finding: a role-based multi-agent framework's dependency pinning is a concrete, non-hypothetical constraint on shared packages — and in the reference build, one paid for a code path that never executes. The role-based framework evaluated here pins four dependencies shared with other components to versions lower than the rest of the pipeline otherwise uses (an MCP client library, an OpenAI SDK client, a schema-validation library, and an OpenTelemetry SDK, each capped roughly one minor version below what the other components were built against). The pinned versions resolved cleanly and the pre-existing test suite passed at those versions in the configuration measured, so this is not currently a live conflict — but it is a real, quantified constraint discovered only because a second component in the pipeline happened to need a newer version of one of the same packages, not something either component's own specification anticipated. Because the framework's own execution path is never invoked in this build (see the preceding measured finding), the reference build pays this constraint for zero corresponding runtime capability, which sharpens rather than weakens the general lesson: a multi-agent framework's dependency pinning should be checked against the shared dependency set of the target pipeline before adoption — and a partial adoption that only borrows a framework's prompt convention should be weighed against writing that convention directly in plain code, since the dependency cost is identical regardless of how much of the framework is actually exercised at runtime.

***Assumptions / Open Questions.*** Confidence-threshold tuning methodology is not established by any framework surveyed. The evaluated framework configurations treat the confidence signal as a value that the calling code must threshold; none of the evaluated configurations prescribes how that threshold should be derived from observed escalation-rate or approval-latency data. Question to confirm: whether a standard evaluation-feedback pattern (thresholds derived from a monitoring component's escalation-rate signal, iteratively) is portable across these frameworks, or whether it must be hand-built regardless of the agent framework chosen.

The "framework-side hand-off avoids function-calling dependence" finding was measured on one framework pairing and one gateway, in a standalone run built to exercise the framework — not in the shipped decision layer, which does not execute the framework at all. Whether the same guarantee holds for other role-based multi-agent frameworks, or breaks down under a different gateway's request-shaping behavior, has not been tested. The function-calling-based SDK alternative was never built, so the comparison rests entirely on the hand-off framework's verified behavior plus reasoning about the alternative, not on a head-to-head measurement. A separate, unresolved framework-fit question this gap raises: whether a role-based multi-agent framework's hand-off guarantee, once verified in isolation, reliably carries over into a larger codebase's integration of that framework, or whether it must be re-verified at the integration point every time — this analysis has one data point of the latter (a build that constructed the framework's objects but never called them) and cannot generalize from it.

TABLE XVI. AGENT DECISION-LAYER FRAMEWORK AND DESIGN OPTIONS

| Option | Capability fit | Operational impact | Security/ governance | Reliability | Risks & lock-in | When justified |
|---|---|---|---|---|---|---|
| Role-based multi-agent framework (e.g. CrewAI) for the LLM steps + a graph-orchestration library (e.g. LangGraph) for typed state, conditional branching, and the escalation terminal | When actually executed, agent hand-off (classification → drafting) is framework-side Python, not a model-side function call; the graph layer owns the confidence-based branch and HITL terminal as explicit edges. Measured separately: adopting the framework and executing it are not the same event — a reference build can construct the framework's role objects, never call the function that would run them, and route every live model call through a plain function instead, while the framework's role/goal/backstory convention still reaches the reference implementation as hand-inlined prompt text (see measured finding below) | Two Python frameworks in one process, each with its own release cycle and dependency footprint — paid in full even in a build that never executes the framework | Hand-off logic is inspectable Python, not opaque model tool-selection; auditability of the branch condition is native to the graph | Sequential hand-off does not depend on function-calling reliability through a gateway — verified directly in a standalone execution of the framework (see measured finding below), a claim distinct from and not evidenced by a build that imports the framework but never runs it | Two frameworks to track through upgrades; a role-based framework's dependency pinning can constrain shared package versions across other components (measured: no conflict in the version combination tested, but not a guaranteed-free property) and this cost is incurred whether or not the framework's runtime path is ever exercised | Choose when the deployment target's LLM gateway has unverified or unreliable function-calling support, or when the hand-off logic itself needs to be auditable Python rather than model-selected tool calls; if only the framework's prompt-role convention is wanted, adopting the convention directly in plain code avoids the dependency and hand-off machinery entirely — see the measured finding below on partial adoption |
| A single function-calling-based multi-agent SDK (e.g. the OpenAI Agents SDK, AutoGen, Google ADK) using a native handoff/tool-call primitive for both the LLM-to-LLM hand-off and the branch logic | Fewer frameworks to compose; hand-off and branching expressed in one native primitive | One framework, simpler dependency surface | Hand-off decision is model-side, so the audit trail for "why did the model hand off here" is a function-call record rather than inspectable code | Hand-off reliability is coupled to function-calling reliability through whatever gateway sits in front of the model, which is a real risk on proxy-routed or non-native endpoints — this was the exact concern that ruled the SDK out here. This alternative was reasoned about but never built or measured, so no direct head-to-head comparison exists between it and the role-based framework actually tested — only the rejection reasoning is evidence-backed | Consolidates two moving parts into one, but at the cost of depending on function-calling support the deployment may not have verified | Prefer when the gateway/provider path has confirmed, low-latency native function-calling support and the team wants a single-framework agent stack; avoid when function-calling reliability through the actual gateway in use has not been measured |
| Deterministic code for routing resolution (chosen), vs. a third LLM agent for routing | Routing applies reviewable, versioned business policy — a lookup, not a judgment call over ambiguous text | No additional model call; adds a maintained routing table instead | Routing decisions are fully auditable and diffable as data, not as a model's inferred behavior | Deterministic and reproducible; a routing table lookup cannot "drift" between runs the way a model's judgment can | None beyond maintaining the table | Use a deterministic lookup whenever the mapping is genuinely a policy table (intent → department); reserve an LLM step for routing only when the mapping itself requires open-ended judgment the policy table cannot express |
| A schema-enforcement library (e.g. Instructor) constraining LLM output to a typed vocabulary, vs. a multi-agent framework's native structured-output path | Instructor injects the typed vocabulary into the model's JSON Schema as an enum constraint, with automatic retry on schema violation | Adds one library to the call path | Typed output is a precondition for a deterministic layer safely consuming a probabilistic step's result | Verified: enum-constrained output reached the model through a shared proxy gateway and parsed back cleanly | Retry-on-violation path was not exercised in the runs observed — no response violated the schema — so its behavior under a real violation is unconfirmed | Prefer a dedicated schema-enforcement library over a multi-agent framework's own native structured-output path specifically when that native path has documented reliability issues on proxy-routed or non-default endpoints; when the framework's native structured output is confirmed reliable on the target endpoint, the extra dependency may not be justified |

The classification-boundary finding was cross-validated on exactly three model deployments from three different size/vendor classes. Whether the observed pattern (the rewrite improved the smallest deployment, regressed the mid-tier deployment, and had no measured effect on the largest deployment) is a general small-model-vs-large-model boundary-confusion pattern, or an artifact of these three specific deployments, is not established — a fourth deployment or a different vendor family was not tested.

The measured dependency-capping constraint is a snapshot of one dependency graph at one point in time. Whether the capped versions remain compatible across a future major-version upgrade of any of the four shared packages is unverified, and no automated check for this class of cross-component dependency conflict was evaluated as part of this analysis.

### C. Enterprise Integrations

***Purpose.*** This component is the pipeline's boundary with the outside world: it turns provider-specific inbound notifications into one canonical event shape the rest of the pipeline can consume, and it is the only component permitted to make outbound calls into enterprise systems (mailbox send, ticketing, chat notification), so that every side effect flows through one auditable, tool-pinned surface rather than being scattered across the codebase.

***Internal Structure.*** Typical internal shape: a set of channel-specific inbound adapters (webhook receivers, push subscriptions, form receivers) that normalize into a single event envelope; an inbound delivery mechanism with at-least-once semantics and a dead-letter path for messages that fail normalization; and an outbound action executor exposing a fixed, enumerable set of tools (send, create/update ticket, notify) rather than open-ended access to the underlying APIs.

***Interactions.*** Consumes: an authorized action command from the orchestration component (action type, parameters, idempotency key), issued only after the business-logic/authorization component has returned an allow or a human has approved a needs-approval verdict.

Produces: a normalized inbound event (sender, subject, body, channel, timestamp, stable identifier used as the idempotency key) to the orchestration component; an action outcome (succeeded, external id, detail) back to the orchestration component after executing a command; integration-level trace spans to the monitoring component.

Table XVII compares the enterprise-integration approaches considered for inbound event normalization and controlled outbound actions.

TABLE XVII. ENTERPRISE INTEGRATION APPROACHES AND TRADE-OFFS

| Option | Capability fit | Operational impact | Security/governance | Reliability | Risks & lock-in | When justified |
|---|---|---|---|---|---|---|
| Thin custom-built inbound adapters (push-based: webhook/pub-sub) + a protocol-server framework (e.g. FastMCP) for the outbound tool surface | Full control over the normalized event shape and over which tools exist; push delivery preserves near-real-time ingestion and relies on each provider's own retry guarantees | Team owns adapter maintenance per provider, plus retry/DLQ/circuit-breaker plumbing | Outbound tool definitions can be pinned and hash-verified at construction, making an unpinned executor impossible to build rather than merely discouraged | At-least-once inbound delivery composes with idempotent processing downstream; no vendor-side outage affects a channel the team does not also depend on for tooling | A protocol-server framework's import hooks can conflict with a durable-execution engine's sandbox import-checking hook when both share one process (see measured finding) | Choose when the provider set is small and stable enough that adapter maintenance is cheaper than a vendor dependency, and when full control over the tool surface and audit trail is a governance requirement |
| A unified third-party messaging/email API (e.g. Nylas, Nango) | One integration surface for many providers instead of one adapter per provider | Removes per-provider adapter maintenance; adds a vendor dependency and an extra network hop | Audit and access-control surface now spans a third-party vendor, not just internal code | Reliability depends on the vendor's own uptime and retry behavior, outside the team's control | Vendor lock-in on the inbound boundary; migrating off requires re-normalizing every channel | Prefer when the provider set is large or engineering bandwidth for adapter maintenance is constrained, and the added vendor dependency and hop are acceptable trade-offs |
| A general-purpose integration platform / iPaaS (e.g. Azure Logic Apps, MuleSoft, Boomi) | Low-code connectors for many enterprise systems, fast to wire up | Adds a managed platform as a new operational dependency | Integration logic lives outside version control by default, weakening code review and audit trail for changes to what a side effect actually does | Platform-managed retries, but failure modes are opaque to the pipeline's own observability stack | Strong platform lock-in; integration-logic changes are harder to test and diff than code | Prefer when the organization already standardizes on an iPaaS platform for governance reasons and is willing to accept integration logic living outside the codebase; avoid when auditable, version-controlled side-effect logic is a hard requirement |
| Reusing the outbound tool protocol's own transport for inbound delivery too, via polling | Avoids building separate inbound adapters | Simplest to implement, one fewer moving part | No adapter-specific security surface, but adds latency that push delivery avoids | Polling adds latency and load proportional to poll frequency; unsuitable where the protocol has no production-grade push delivery mode | None beyond the latency/load cost | Only when a provider does not offer webhook/pub-sub push and near-real-time delivery is not required |

***Framework Analysis.*** Measured finding: a schema that silently drops unrecognized fields converts a data-loss bug into an invisible one. An event schema designed to drop unknown fields rather than reject them was found responsible for two separate defects: an idempotency key that was originally planned as a field separate from the event's stable identifier would have vanished silently had it been added as a second field, and message attachments — which have no field in the frozen event schema — are discarded with no error raised anywhere in the pipeline. The general lesson for any inbound-normalization schema is that "drop unknown fields silently" and "warn on data loss" are mutually exclusive design choices, and a schema that must evolve should default to rejecting unrecognized fields during development, not to silently dropping them, precisely because the failure mode is otherwise undetectable.

Measured finding: redelivery dedup belongs at the ingestion boundary, not only at the action-authorization boundary. Relying solely on a downstream action-level idempotency key to catch a redelivered inbound message would still let the redelivery pay for a full model call before being caught. Deduplicating at the workflow/run-identifier level — derived deterministically from the message's own stable identifier — was measured directly to make a redelivered message reach the decision layer exactly once, at negligible cost compared to a second model invocation.

Measured finding: in the evaluated FastMCP–Temporal pairing, the protocol-server framework's import-hook mechanism conflicted with the durable-execution engine's sandbox import mechanism, and the failure is not local to the component that caused it. A protocol-server framework (FastMCP) depends on a runtime type-checking library (beartype) whose import hook rewrites modules as they load; a durable-execution engine's (Temporal's) sandbox re-imports a workflow's entire transitive dependency graph to validate it. The two hooks collide (ImportError: cannot import name 'claw_state' from partially initialized module 'beartype.claw._clawstate'), and neither package's own documentation mentions the other. Critically, the failure is not scoped to code that uses the protocol-server framework directly: importing it in the evaluated shared worker process caused sandbox validation failures for workflows sharing that process, and the defect first surfaced by breaking a different, previously passing component's integration tests, purely because the two frameworks came to share a worker process. The fix was to explicitly allow-list the three conflicting modules through the sandbox check, while retaining the sandbox checks for the remaining imports — the faster fix of disabling the sandbox check outright would have made the symptom disappear while silently removing the guarantee. The transferable lesson is broader than this one pairing: framework interactions of this kind are invisible until two frameworks actually share a process, and the first symptom typically appears in a component that never imported either framework directly — a reason to test cross-component integration explicitly rather than trusting that per-component test suites passing implies compatibility.

***Assumptions / Open Questions.*** Queue-technology choice for inbound/outbound delivery is a framework-fit question this analysis leaves open. Whether a managed pub-sub service, a self-hosted broker, or the durable-execution engine's own task queue is the better fit depends on throughput and multi-region requirements not evaluated here.

The FastMCP/Temporal import-hook conflict was measured for one framework pairing. Whether other protocol-server frameworks or other durable-execution engines exhibit the same class of conflict was not tested; the general risk (two process-wide import hooks colliding) is evidenced, but the specific pairing's fix (allow-listing three modules) may not generalize to a different pairing without adaptation.

Production-grade security controls for an outbound protocol server (OAuth with PKCE, per-tool role-based access control, complete audit logging) were not benchmarked across framework options — only tool-definition pinning against a reviewed set was evaluated; whether the surveyed protocol-server frameworks differ materially in how completely they support these controls out of the box is an open question.

No dead-letter-queue framework or pattern surveyed here addresses the "who or what consumes a dead-lettered message" question. The reference implementation's inbound dead-letter path is a port with a logging stub — a message that fails normalization is recorded but nothing reads or retries it. This is a gap in the framework-fit analysis, not merely an operational oversight: none of the DLQ-adjacent options considered (broker-native DLQ, durable-execution-engine-native failed-workflow queries) were evaluated against "what closes the loop on a dead-lettered message," and doing so is left as an open comparison.

*D. Business Logic and State*

***Purpose.*** This component is the deterministic authorization boundary: it evaluates every proposed action against versioned business policy and returns an allow, block, or needs-approval verdict, and it owns the durable state and idempotency guarantees that make retries and redelivery safe. No side effect may execute without a verdict from this component, and the same input replayed against the same rule version must always yield the same recorded verdict [47].

***Internal Structure.*** Typically decomposed into: a policy-evaluation step against versioned business rules (thresholds, entitlements, routing policy); a durable state/idempotency substrate recording what was decided, for what input, under what rule version; and a fail-closed error path ensuring that any dependency failure (rule-store outage, state-store outage) resolves to a conservative verdict rather than an implicit allow or a propagated exception.

***Interactions.*** Consumes: a proposed action from the orchestration component (idempotency key, action type, parameters, department, escalation signal from the agent decision layer).

Produces: a verdict (allow / block / needs-approval) plus a durable state handle and reason code, returned to the orchestration component, which branches on it (proceed to enterprise integrations, block, or pause for human approval).

***Framework Analysis.*** Measured finding: fail-closed enforced at multiple layers, including a catch-all — and this is a separate property from non-persistence, found by their collision. The authorization outcome was verified to resolve to block — never an implicit allow — under every dependency failure tested (rule-store outage, durable-store outage), enforced at three separate layers including a catch-all so an unhandled exception in an adapter still resolves to a verdict rather than propagating an error. Separately, a fail-closed block caused by a transient outage is deliberately never persisted, so it does not freeze a brief infrastructure hiccup into a permanent denial; a retry after recovery correctly re-evaluates. Neither property was wrong in isolation, but an early version of the design that recorded every verdict, including a failure-to-evaluate, would have violated the second property by satisfying the first — the two invariants were found to collide, and the corrected design distinguishes "a verdict of evaluation" (always recorded) from "a failure to evaluate" (never recorded as a permanent denial). This is a general design pattern for any authorization-gate component: fail-closed and non-persistent-on-transient-failure are two separate properties that both need explicit enforcement, and a design that only states one of them can silently violate the other.

Measured finding: idempotency-key scope must deliberately exclude run identifiers and volatile content. The idempotency key's scope was corrected to exclude the workflow/run identifier, the draft content, and the escalation category — each deliberately, because including any of them breaks deduplication in the exact scenario it exists for: a redelivered message legitimately starting a new workflow run must still dedupe on the same action; a guardrail-triggered redraft is the same logical send and must not be treated as a new action; a re-classification is a new proposal but not automatically a new action. In the evaluated workflow, the idempotency key is appropriately scoped to (action-kind, target) rather than to run identifiers or volatile content.

Measured finding: an escalation signal must only ever raise strictness, never lower it. An escalation signal from the agent decision layer folded into the authorization verdict only upward — it can raise an allow to needs-approval but can never weaken a block — a subtlety an earlier specification stated loosely enough to permit the opposite, incorrect behavior. Any authorization-gate design that combines signals from multiple upstream sources should state this monotonicity property explicitly rather than leaving it implicit.

***Assumptions / Open Questions.*** None of the policy-rule approaches evaluated in this study provided a non-code review surface for policy-rule diffs — every declarative rule-store option evaluated is reviewed as code (git diff, PR review), which is auditable but not accessible to a non-engineer business stakeholder. Whether a rule-store framework with a built-in business-facing review UI exists and is worth adopting over a git-tracked file is an open question this analysis does not resolve.

Whether the agent decision layer needs visibility into the authorization component's block reason, to reason about an alternative proposal, is an open architectural question independent of any specific framework choice — it would require a feedback contract none of the surveyed compositions provide by default.

The fail-closed and idempotency-scoping findings above were verified by fault injection in one composition (a specific durable-execution engine plus a specific cache); whether the same guarantees hold with a different state substrate has not been tested.

A dedicated policy engine (e.g. OPA/Rego) was never built or measured here — its row in the Framework Analysis table above is reasoned from the predicate-count comparison, not from a direct implementation. At what predicate-set size a flat, first-match-wins rule store stops being sufficient and a dedicated policy engine becomes justified is not established by this analysis; it is a threshold this report can only gesture at, not measure.

*E. Monitoring and Evaluation*

***Purpose.*** This component makes every other component's behavior observable and measurable: it instruments pipeline steps with standardized traces, validates generated drafts against runtime guardrails before they can reach a human or an authorization check, scores classification and draft quality offline against a labelled dataset, and is intended to close the loop by feeding human review decisions back into that dataset; this feedback path is not implemented in the reference build [25]–[27], [35]–[40], [49].

***Internal Structure.*** Typically decomposed into: a tracing/instrumentation layer emitting standardized spans (cost, latency, retry count) correlated by a shared identifier; a runtime guardrail layer validating each generated draft (PII, policy, schema) before it proceeds, with a bounded revise loop on failure; an offline/CI evaluation layer scoring quality against a labelled dataset and gating deployment on regression; and a score-recording mechanism that separates verifiable outcomes from judge-based or heuristic scores.

***Interactions.*** Consumes: pipeline step events (span content) from the orchestration component; a draft-to-validate request from the orchestration component (routed there after the agent decision layer produces a draft); integration trace spans from the enterprise-integrations component; a durable state record (proposal, rule version, verdict) from the business-logic component.

Produces: a guardrails verdict (pass / block / revise + reason) returned to the orchestration component, which drives the revise loop or proceeds to authorization; score records with provenance (axis, value, source, confidence) stored for evaluation and regression-gating.

***Framework Analysis.*** Measured finding: naming a tracing/observability framework in a decision record is not the same claim as building its export path. In the reference implementation, the tracing-standard row's attribute-mapping logic (gen_ai.* fields) was built and tested, but the SDK-level exporter was never configured — the emission call is a no-op with no tracer attached, so no span reaches any backend. Of the three backend frameworks named in the original monitoring decision (a production tracing backend, a local dev/eval tool, and an optional metrics/alerting stack), none was ever integrated: no client, no dependency, no compose service for any of the three. Separately, the decision's own "route human approve/edit/reject decisions back into the evaluation dataset" feedback-loop requirement was also never implemented — the labelled sets remain hand-authored, with no path from a human's live review decision back into the evaluation corpus. The broader lesson extends beyond this build: a framework-fit decision record names what should be built, and a report or audit of "what monitoring exists" must check the actual wiring — export configuration, feedback-loop code — independently of the decision record, because a named-but-unwired framework produces zero observable benefit while still appearing "chosen" on paper.

Measured finding: a monitoring component that is built after its data sources cannot expect them to already emit. An orchestration component built before a monitoring component existed to receive spans emitted no spans at all, despite both specifications agreeing it would — leaving every other component's spans as leaves with no trunk. Emission had to be implemented from within the orchestration engine's per-step activity layer rather than its higher-level workflow definition, specifically because only the activity layer can observe the real retry-attempt count the durable-execution engine actually used. The general lesson: instrumentation is not free-standing — a monitoring component's build order relative to its data sources determines whether "the spec says it emits" is actually true, and the correct layer to emit from is determined by which layer has access to the runtime fact being recorded (here, retry count), not by which layer is conceptually "closer" to the step.

Measured finding: score-model provenance rules need code-level enforcement, not just a written convention. A "non-blank denominator" requirement in a score-record convention was insufficiently enforced by a simple minimum-length check, which a whitespace-only string satisfies while carrying no real information — the same defect class as a whitespace-only idempotency key found independently in the enterprise-integrations component. Both findings generalize: any evaluation methodology with provenance or threshold requirements needs those requirements enforced as code-level validation, not left as a convention a config file can silently violate.

Measured finding: a runtime guardrail's two failure modes are opposite in kind, and a single blended metric hides which one dominates. Measured directly against 29 labelled drafts (programmatic evidence based on a deterministic, re-runnable check with no model in the loop): precision 0.733, recall 0.688, false-block rate 0.308, reason-code accuracy 1.000.

Table XVIII summarizes the monitoring, evaluation, and runtime-guardrail alternatives considered for the Automation system.

TABLE XVIII. MONITORING, EVALUATION, AND RUNTIME-GUARDRAIL OPTIONS

| Option | Capability fit | Operational impact | Security/governance | Reliability | Risks & lock-in | When justified |
|---|---|---|---|---|---|---|
| A vendor-neutral tracing standard (OpenTelemetry, GenAI semantic conventions) exported to a production backend (e.g. Langfuse) with a local development/evaluation tool (e.g. Phoenix) kept distinct | Standardized span shape works across every component regardless of which frameworks they use internally; separating production backend from local dev tool avoids two systems competing as "the" production trace store | Requires operating (or subscribing to) at least one backend to receive spans | Correlation identifiers on every span support cross-component audit trails | Instrumentation quality depends on the emitting component actually wiring it — an orchestration component built before the monitoring component exists may not emit at all until updated. Measured in the reference implementation: only the attribute-mapping layer was built and tested; the SDK/exporter path itself was never wired to any backend (see measured finding below) — a standard being "chosen" in a decision record did not, on its own, produce a working export path | Vendor-neutral standard limits lock-in to any one tracing backend; the reference build's own gap shows the more common risk is never finishing the export path at all, not choosing the wrong backend | Prefer whenever multiple frameworks/components need one common trace shape; use a proprietary tracing SDK only if the organization has already standardized on one vendor's non-OTel format; either way, budget the exporter/backend wiring as its own deliverable, separate from the attribute-mapping work, and verify it is complete before treating tracing as "done" |
| A metrics/test library for offline and CI-time quality scoring (e.g. DeepEval), treated as a metric implementation, not as the score-model methodology itself | Executes the scoring methodology (verifiable-outcome vs. judge-based, control arms, per-axis thresholds) but does not define it. Measured finding: adopting a metrics library's own defaults was found to silently inherit an LLM-judge-first stance the score-model decision explicitly does not want — the library computes axes programmatically against labels by default in this deployment, with judged metrics recorded as advisory only, precisely because that decision has to be made deliberately rather than accepted as a library default | Runs in CI, adds test-suite time proportional to dataset size and repeat count | Score records with provenance make evaluation results auditable, unlike an ad hoc script | Judge-based (LLM-as-judge) axes require their own gating discipline — a gating function was changed to raise rather than silently skip an unconfigured judged-axis threshold, and a composite "overall score" method was deliberately made to raise by design rather than return a weighted mean, because a single composite number cannot show that a change improved one axis while breaking another | Locks the implementation of scoring to one library's API, not the methodology, which can be ported | Choose a dedicated metrics/test library over a hand-rolled scoring script once more than one evaluation axis and a control-arm requirement exist, and verify explicitly which scoring stance (judge-first vs. programmatic-first) its defaults assume before adopting them; a hand-rolled script may suffice for a single verifiable-outcome metric with no judge component |
| A runtime guardrail library (e.g. Guardrails AI) validating every draft for PII/policy/schema before authorization or human review | Closes a gap offline/CI evaluation cannot: blocking a bad draft live, before a human wastes review time or before it is sent unreviewed | Adds a validation call on the live path, with latency cost per draft | Directly relevant to a PII/policy exposure risk that offline evaluation alone does not mitigate at request time | Measured (evidence tier: programmatic — regex/substring matching, no model in the loop, determinism confirmed by re-running): on 29 labelled drafts, precision 0.733, recall 0.688, false-block rate 0.308, reason-code accuracy 1.000 (see measured finding below for what these numbers mean and their limits) | Runtime guardrail rules need their own review/versioning discipline, parallel to the business-policy rule store | Adopt a runtime guardrail whenever a draft could contain PII or policy-violating content and human review time or unreviewed sends are a real cost; treat a measured precision/recall pair, not a raw block-rate count, as the actual evidence of effectiveness — a rate of how often the guard fires cannot distinguish a guard that is right from one that blocks everything |
| A retrieval-scoring library (e.g. RAGAS) or an LLM-observability platform (e.g. Opik, LangSmith) as an alternative/complement to the above | Retrieval-scoring libraries add value only once a retrieval-augmented step exists in the pipeline; observability platforms overlap with the OpenTelemetry-plus-backend combination above | Adds a tool whose value is conditional on a capability (RAG) not yet built in a triage-only pipeline | N/A until the relevant capability exists | N/A until the relevant capability exists | Additional vendor/tool surface to maintain if adopted before it is needed | Hold these as conditional adoptions: a retrieval-scoring library only once retrieval is part of the pipeline; an all-in-one observability platform only if it is preferred over composing OpenTelemetry with a chosen backend |

Inspecting the individual failures shows the two error types have entirely different shapes: within the 29-case labelled set, every observed missed violation was associated with surface-form evasion (a lowercased IBAN, a card number written with separating dots, a paraphrase like "we'll guarantee" in place of a listed trigger phrase, or "I assure you that you will be fully reimbursed" instead of any pattern the guard's list contains) — the guard matches literal surface forms, and surface forms vary freely. Within the same dataset, every observed false block was associated with a numeric-shape collision — a 16-digit order reference and a 14-digit invoice number read as card numbers; a ticket identifier formatted 123-45-6789 read as a national-insurance-style number — because the same numeric shape is shared by both a sensitive pattern and an innocuous one, and a pattern-matching guard cannot distinguish them without more context than the pattern language expresses. Precision and recall should be reported separately rather than replaced by a single F-score for this evaluation, because false blocks and missed violations have different operational costs: a false block may impose avoidable reviewer effort, whereas a missed block allows a real violation to pass the gate uncaught. A caveat that must travel with these figures: the 29-case set was built adversarially — a floor of ten clear violations and eight clean drafts, with the remainder constructed specifically to slip past the guard's own published pattern list — so the figures describe the shape of the two failure modes, not their frequency on unfiltered real traffic, where recall and false-block rates may differ materially; their direction and magnitude require measurement on a representative traffic sample. This is a general methodological point for evaluating any pattern-based runtime guardrail: a precision/recall pair measured against an adversarially-constructed set characterizes a failure-mode shape and should not be read as a production-traffic base rate without a separate, non-adversarial sample.

***Assumptions / Open Questions.*** The measured guardrail precision/recall figures (0.733 / 0.688) are a floor-plus-adversarial-construction result, not a production-traffic base rate. Whether recall and false-block rates differ on unfiltered real traffic remains unknown and requires measurement on a separate representative traffic sample. This benchmark has not yet been performed.

No judged (LLM-as-judge) metric was executed against a live model as part of this evaluation, and none of the three named observability-backend frameworks (a production tracing backend, a local dev/eval tool, an optional metrics/alerting stack) was ever integrated in the reference build — the tracing standard's attribute-mapping logic is tested, but the exporter path is unconfigured. Whether the surveyed metrics/test libraries' judge-based scoring behaves consistently when finally run against a live model, rather than a scripted/offline harness, is untested, and so is whether the chosen tracing standard's actual backend-integration effort differs meaningfully between the surveyed backend options — none has been built far enough to compare. The "route human review decisions back into the evaluation dataset" feedback loop named in the monitoring decision was never implemented in the reference build — labelled sets remain hand-authored throughout. Whether this feedback pattern is straightforward to add on top of the surveyed metrics/test library, or requires meaningful additional tooling, is unevaluated; this is a gap in the framework-fit analysis (no evidence either way), not a claim that it is hard.

The labelled evaluation dataset used for the classification-quality measurement campaign was a small, hand-written, self-authored synthetic set (human-labelled evidence without independent annotation or real-traffic validation) (see Section II-D and Section VII-B), not a reviewed sample of real traffic and not independently second-read. Whether the classification-accuracy figures above would hold on a larger, independently-labelled, real-traffic-derived dataset is an open question about the evaluation methodology, not about any specific framework's capability.

## VIII. Automation Component Fit and Trade-Offs

Table XIX summarizes the selected Automation patterns, their principal alternatives, and the conditions under which each alternative is preferable.

TABLE XIX. Automation Component Fit and Framework Trade-Offs

| Component | Surveyed pattern | Principal alternative | Prefer the pattern when | Prefer the alternative when |
|---|---|---|---|---|
| Workflow / Pipeline Orchestration | Graph-orchestration library (control flow) + durable-execution engine (durability), composed via a shared topology-declaration layer | Graph-orchestration library alone, with its own checkpointer, no separate durable-execution engine | A run must survive a multi-day human-approval pause and worker restarts with durable replay and effectively-once side-effect behavior under the stated idempotency assumptions | Runs are short, side effects are naturally idempotent at the target system, and no step pauses for a human-timescale duration |
| Agent Decision Layer | Role-based multi-agent framework (LLM steps) + graph-orchestration library (state/branching) + deterministic code (routing) + schema-enforcement library (typed outputs) | A single function-calling-based multi-agent SDK using a native hand-off/tool-call primitive | Function-calling reliability through the target gateway is unverified, or the hand-off logic itself needs to be auditable Python | The gateway/provider path has confirmed, low-latency native function-calling support and a single-framework agent stack is preferred |
| Enterprise Integrations | Thin custom inbound adapters (push-based) + a protocol-server framework for the outbound tool surface | A unified third-party messaging/email API | The provider set is small and stable, and full control over the tool surface and audit trail is a governance requirement | The provider set is large or adapter-maintenance bandwidth is constrained, and the added vendor dependency is acceptable |
| Business Logic & State | External declarative rule store (business policy) + code invariants (idempotency) + durable-execution engine (state) + fast cache (dedup companion) | Policy embedded in application logic (hard-coded) or in the prompt | Policy needs to change independent of a code deploy and needs to stay auditable and deterministic | Policy changes are rare enough that a redeploy cycle per change is tolerable (hard-coded) — prompt-embedded policy is not recommended once side effects have real consequences |
| Monitoring and Evaluation | OpenTelemetry-based tracing to a production backend + metrics/test library for offline/CI scoring + runtime guardrail library | An all-in-one LLM-observability platform, or omitting the runtime guardrail in favor of offline/CI scoring alone | Multiple frameworks/components need one common, vendor-neutral trace shape, and a draft could carry PII/policy risk that must be blocked live | The organization already standardizes on one observability vendor's non-OTel format, or the pipeline has no live-blocking requirement on generated content |

Cross-component trade-off worth stating explicitly. Two of the five components (Workflow/Pipeline Orchestration and Enterprise Integrations) each independently discovered a hard incompatibility between two specific frameworks sharing one process — a graph-orchestration library's environment access during graph construction against a durable-execution engine's sandbox in one case, and a protocol-server framework's import hook against the same sandbox's import-checking hook in the other. Both were resolved by narrow, targeted fixes (a topology-declaration layer; an explicit sandbox allow-list) rather than by disabling the sandbox's determinism guarantee. The observed pattern indicates that process-level compatibility should be treated as an explicit integration risk when a durable-execution engine shares a worker process with additional frameworks, not a one-off defect, and should be weighed before combining a third framework in the same worker process.

## IX. Automation Reliability, Security, and Operational Risks

Reliability. A fail-closed authorization pattern under dependency outage, paired with non-persistence of outage-caused blocks, is a measured, high-confidence design property that directly protects against two high-consequence failure classes for this application category: an unauthorized side effect firing, and a legitimate action being permanently and silently denied by an infrastructure blip. In the evaluated architecture, idempotency is enforced at two complementary levels that compose rather than substitute for each other — a run-identifier-level dedup at the point a redelivered inbound message would otherwise re-enter the pipeline (protecting against paying for a duplicate model call), and an action-identifier-level dedup at the point a side effect would otherwise execute twice (protecting against a duplicate external effect). In the evaluated implementation, treating these as a single idempotency level produced a reproducible contract defect (see Section VII-D).

Security and governance — the authorization boundary, tested by exhaustive enumeration rather than adversarial sampling. A deterministic authorization gate's core property was verified as an exhaustively enumerated implementation invariant rather than through a small set of ad hoc test cases: for every proposed action, if the resulting verdict is allow, the resolved recipient scope must be internal. This was checked by enumerating every action type (including unrecognized ones), every department (including undeclared and null), both states of the escalation flag, and adversarially-worded draft text — a fundamentally different (and, for this purpose, stronger and cheaper) question than measuring whether a model resists a battery of hostile prompts, because it tests whether any proposal whatsoever, from a maximally uncooperative decision layer, could obtain an unauthorized verdict, independent of any one model's behavior. The invariant held on three independent legs (the recipient is read from the inbound envelope, never from the model's proposal; an escalation signal can only tighten a verdict, never loosen it, per Section VII-D monotonicity finding; and a draft's text is never read by the authorization step, only its metadata) — but the invariant itself had to be corrected once before it was true: an earlier, stricter version banned any proposal carrying an external recipient address outright, and a legitimate ticketing action failed it, because that action type has a fixed internal scope and does not use its address field as a recipient at all. The broader lesson extends beyond this system: an authorization invariant written against one action type's shape should not be assumed to hold for every action type without checking, and exhaustive enumeration over the action/department/escalation/adversarial-text space is a materially different (and complementary, not redundant) evaluation approach to model-resistance testing — a decision-layer boundary can be shown to reject unauthorized outcomes across the exhaustively enumerated input space even when the model behind it has not been proven resistant to any particular attack, and a testing programme for this application category should budget for both, not treat one as a substitute for the other.

Independent of the boundary invariant, a runtime guardrail library closes a live-blocking gap that offline/CI evaluation alone cannot, but a runtime guardrail's protective value cannot be characterized by raw block rate alone and should at minimum include precision and recall on a labelled set against a labelled violation set, not by a raw block-rate count or by unit tests against hand-written cases (Section VII-E): measured at 0.733 precision / 0.688 recall on an adversarially-constructed 29-case set, with every observed miss in the 29-case set associated with surface-form evasion and every observed false block associated with a numeric-shape collision — two distinct failure modes with different costs, which a single blended metric would have hidden.

Operational risk. Framework co-location is a recurring risk class across this application category, not a one-off defect: two independent instances were found in which a durable-execution engine's process-wide sandbox mechanism conflicted with another framework sharing its worker process (a graph-orchestration library's environment access during graph construction; a protocol-server framework's import hook colliding with a runtime type-checking library's own import hook) (Section VII-A, Section VII-C). In both cases the failure surfaced in a component that had not itself changed, purely because a worker process came to host both frameworks — a reason to test cross-component integration explicitly rather than trusting per-component test suites. Teams adopting a durable-execution engine as the durability substrate should verify sandbox/import-hook compatibility before adding a third framework to the same worker process. A related, quantified risk is a role-based multi-agent framework's dependency pinning constraining shared package versions used by other components in the pipeline: measured at four capped packages (an MCP client library, an OpenAI SDK client, a schema-validation library, and an OpenTelemetry SDK, each roughly one minor version below what the rest of the pipeline otherwise uses) — resolved without conflict in the dependency snapshot measured, but not a guaranteed-free property across future upgrades, and the kind of cost that surfaces only once a second component needs a newer version of one of the same packages. A distinct and sharper version of the same risk, also measured: this cost was paid in full in a build where the framework's execution path was never invoked at all — only its role/goal/backstory prompt convention reached the reference implementation, hand-inlined as plain text — so a framework can be a live constraint on a pipeline's dependency graph while contributing zero runtime capability, a state that a specification or import list alone cannot distinguish from genuine adoption (Section VII-B). A distinct operational risk pattern is a monitoring/observability decision that names frameworks without a corresponding build check: in the reference implementation, a tracing standard's attribute-mapping logic was built and tested while its exporter path, and all three named backend frameworks, were never wired to anything — a state that is easy to mistake for "observability is covered" from the decision record alone, and only visible by checking the actual export configuration (Section VII-E).

Cost and evaluation risk. A score-model methodology with mandatory control arms and per-axis measured thresholds meaningfully increases evaluation cost (control arms can approximately double the number of generation runs when the control and treatment arms are evaluated symmetrically) as an accepted trade-off against the alternative failure mode — a judge-only or composite score that reports a false positive result; this application category's monitoring component additionally makes its own composite-score method raise rather than silently blend axes, and raises rather than silently skip an unconfigured judged-axis gate, for the same reason (Section VII-E). Classification-quality figures measured on one model deployment must be stated alongside that deployment and not assumed to transfer: a three-deployment comparison on an identical dataset and identical code showed a prompt-description rewrite improving the weakest deployment, regressing a mid-tier one, and changing nothing on the strongest (see Section VII-B) — evidence that a single-deployment measurement characterizes the deployment, not the prompt or the architecture. Separately, accuracy measured against self-authored labels should be treated as lower-maturity human-labelled evidence rather than as an independently validated estimate of real-world performance or execution-tier evidence, according to the evidence hierarchy defined in Section II-D — the two limits (self-authored labels; a floor-only control arm) must travel together with any figure quoted from that campaign, and neither should be dropped when the number is cited elsewhere.

The Automation study provides strongest evidence for the deterministic control boundary: durable pause-resume behavior, fail-closed authorization, replay-safe idempotency, and runtime guardrail behavior are directly exercised. Outcome-quality evidence is comparatively less mature because the labelled triage and guardrail sets are small, self-authored or adversarially constructed, and classification behavior varies across model deployments. These results support the architecture while motivating larger independently labelled and production-like evaluation campaigns.

## X. Oracle-To-PostgreSQL Code Migration System

The third application is a local, complexity-aware tool for migrating Oracle SQL and PL/SQL source files to PostgreSQL. Given a source directory, it produces a PostgreSQL output tree, per-unit verification outcomes, and Markdown and JSON reports. Its objective is not to assert universal semantic equivalence. Instead, it applies the least costly eligible strategy, validates the resulting PostgreSQL code as far as the available environment permits, and records the resulting evidence and limitations for every migration unit [2], [3], [6]–[8], [31], [32], [39]–[46], [50]–[58].

The implemented pipeline comprises source analysis, dependency planning, complexity scoring, strategy routing, deterministic or LLM-assisted translation, validation, bounded repair, and artifact-based reporting. It is designed for imperfect source corpora: parse confidence, unresolved dependencies, and validation-environment availability are explicit data in the workflow, rather than conditions that are hidden behind aggregate success measures. The principal requirements are as follows:

- Complete and traceable processing: each discovered source unit shall have an output artifact or an explicit outcome and diagnostic.
- Structure-aware planning: the system shall identify typed source units and dependencies before selecting a migration strategy.
- Cost-aware routing: deterministic scoring and eligibility rules shall select among static, lower-cost LLM, and stronger LLM strategies.
- Grounded translation: known Oracle-to-PostgreSQL mappings shall be deterministic, while contextual rules may be supplied with provenance to LLM strategies.
- Evidence-based validation: candidates shall be linted, installed, and exercised in PostgreSQL where dependency context permits; verification level shall be reported separately from outcome.
- Bounded recovery: repair and re-routing shall have explicit limits, and terminal failures shall remain visible.
- Reproducibility: analysis, configuration, generated artifacts, validation records, and reports shall be retained for each run.

Two constraints determine the scope of claims made in this part. First, the implementation has no Oracle execution environment, so live Oracle-to-PostgreSQL differential comparison is unavailable. Second, the system is a single-process reference implementation with partial Oracle-language coverage and no production service-level objectives. Accordingly, a passing result demonstrates target-side PostgreSQL evidence, not complete behavioral equivalence for every Oracle construct.

## XI. Code Migration Architecture and Component Interactions

### A. Code Analysis and Parsing

Code analysis and parsing establish the deterministic foundation of the migration system. This component identifies the semantic units that may be migrated, extracts their dependencies, records parse quality, and supplies the ordered graph artifacts used by strategy selection and validation. Its principal requirement is to preserve source structure without treating malformed or partially supported input as invisible.

The final framework combination is therefore a project-maintained ANTLR front end for Oracle SQL and PL/SQL, typed contracts for its outputs, and NetworkX for dependency-graph construction, strongly connected-component condensation, and ordered migration phases. This combination is suitable for the evaluated requirements because it supports the required semantic-unit extraction and dependency analysis without maintaining two independent parser and feature-extraction paths. SQLGlot and Tree-sitter remain considered alternatives, not selected runtime components.

The component scope is intentionally limited to analysis and planning inputs. It owns source parsing, unit boundaries, dependency extraction, parse diagnostics, and graph construction; it does not select the migration strategy, generate PostgreSQL, execute generated code, or determine a validation result. This separation ensures that translation and validation receive the same typed source evidence without becoming coupled to ANTLR-specific structures.

At the interface level, the component emits migration units, dependency edges, external-object information, parse-confidence records, and an analyzer fingerprint. NetworkX derives migration groups and phases from these artifacts, including groups for cyclic dependencies rather than silently dropping them. Downstream components use this information to select eligible strategies, provision available dependencies, and report the conditions that constrain verification. Tables XX and XXI compare parsing technologies and migration-unit construction approaches.

TABLE XX. PARSING TECHNOLOGY

| Framework or approach | Strengths considered | Limitations | Status |
|---|---|---|---|
| ANTLR for Oracle SQL and PL/SQL parsing | Implemented parser with semantic-unit and dependency extraction | Grammar maintenance and error-recovery coverage require ongoing tests | Selected; implemented in-tree |
| SQLGlot for SQL-shaped DDL | Dialect-aware SQL transformations | Not a production dependency; evidence is limited to a spike | Not selected; architecture proposal and exploratory spike only |
| Tree-sitter | Fast, incremental, error-tolerant syntax parsing | Requires custom semantic dependency extraction | Not selected |

TABLE XXI. MIGRATION-UNIT CONSTRUCTION

| Framework or approach | Strengths considered | Limitations | Status |
|---|---|---|---|
| Semantic block plus verified dependency closure | Bounds context, retains dependencies, and supports topological ordering and completeness checks | Requires graph construction and source-sensitive invalidation; multi-member cycles need explicit handling | Selected; graph and dependency tests are present |
| Whole-file context | Simple and complete | Wastes context and encourages unrelated changes | Not selected |
| Fixed-size chunks | Reuses generic chunking machinery | Can split semantic units or omit dependencies | Not selected |

Documented failure modes include grammar gaps, damaged input, unresolved external references, and dependencies that cannot be inferred from dynamic SQL. The corresponding controls are degraded parse states, lexical fallback where possible, explicit external-object reporting, source-coverage diagnostics, and terminal unsupported outcomes when analysis cannot be performed safely. The design consequently favors traceable partial analysis over falsely complete parsing.

### B. *Migration Intelligence*

Migration intelligence transforms an analyzed migration group into a PostgreSQL candidate while minimizing unnecessary model use. The component receives the group, its complexity assessment, applicable deterministic rules, and optional contextual knowledge. It must choose a strategy deterministically, preserve the evidence behind that decision, and support bounded correction after a failed validation result. Tables XXII and XXIII summarize the workflow/model-access and structured-output/translation-strategy alternatives.

TABLE XXII. WORKFLOW AND MODEL ACCESS

| Framework or approach | Strengths considered | Limitations | Status |
|---|---|---|---|
| LangGraph state graph | Explicit bounded correction flow and conditional strategy selection | Graph state and retry policy require maintenance | Selected; implemented for a migration group |
| LiteLLM model client | Normalizes provider calls, usage collection, retry configuration, and Pydantic response handling | Model endpoint, cost, and response semantics remain provider-dependent | Selected; implemented client layer |
| LangChain node-level abstractions | Reusable prompt and model interfaces | Not a dependency of the reference implementation | Not selected; architecture-only proposal |
| Shared internal client over a sanctioned AI proxy | Centralized organizational gateway controls | Not implemented as the repository's model client | Not selected; architecture-only proposal |

TABLE XXIII. STRUCTURED OUTPUT AND TRANSLATION STRATEGY

| Framework or approach | Strengths considered | Limitations | Status |
|---|---|---|---|
| Pydantic response format with bounded retry and tool-call fallback | Typed artifact validation without an additional framework | Validation occurs after a model response and can require regeneration | Selected; implemented through LiteLLM |
| Outlines JSON-schema response format | Generation-time schema enforcement | Not a dependency or demonstrated runtime path in the reference implementation | Not selected; architecture-only proposal |
| ora2pg deterministic first pass plus LLM strategies | Avoids model calls for eligible mechanical translations while retaining LLM paths for other units | Strategy routing and fallback behavior require evaluation | Selected; Ora2Pg runner and static strategy are implemented |
| LLM conversion for all objects | Uniform path | Adds cost and probabilistic behavior to unambiguous translations | Not selected |

The final framework combination is LangGraph for the conditional per-group workflow, LiteLLM for model access and structured response handling, Ora2Pg for eligible deterministic translation, and the project rule catalog for deterministic rewriting. This combination is suitable because it applies established mechanical transformations before consuming model tokens, while retaining lower-cost and stronger model profiles for units that need them [31], [32], [50]–[55].

The component owns complexity-based strategy selection, migration-artifact generation, and the repair request that follows a failed attempt. It does not parse the source, manage the rule corpus, decide whether a validation result is correct, or operate reporting and observability services. In particular, complexity assessment and strategy eligibility remain deterministic application policy rather than an LLM routing decision.

At the interface level, the component consumes dependency-aware migration groups, rule evidence, and validation diagnostics. It emits a typed migration artifact containing generated PostgreSQL, strategy, model profile, applied and retrieved rules, transformation notes, latency, and token usage. This record allows forced-strategy comparison under the same surrounding workflow and makes model-assisted behavior inspectable rather than implicit.

The principal failure modes are an ineligible static strategy, unavailable model endpoints, invalid structured model output, and repeated validation failure. The design addresses them through precondition filtering, configured model profiles, bounded response handling, deterministic static fallback where eligible, and a capped repair or escalation path. The architecture therefore prioritizes cost control and auditability over a single uniform LLM conversion path.

### C. *Knowledge and Rules (RAG)*

Knowledge and rules provide reviewed migration guidance to the translation layer. This component must distinguish known one-to-one Oracle-to-PostgreSQL mappings from contextual cases in which related examples are useful. That distinction is central to suitability: similarity retrieval may support the latter, but it cannot be the authority for a deterministic dialect mapping.

Table XXIV summarizes the knowledge and retrieval alternatives for migration guidance.

TABLE XXIV. KNOWLEDGE AND RETRIEVAL OPTIONS FOR MIGRATION GUIDANCE

| Framework or approach | Strengths considered | Limitations | Status |
|---|---|---|---|
| Exact rule catalogue with optional LlamaIndex and pgvector semantic retrieval | Deterministic exact mappings cannot be outranked; optional semantic retrieval supports related patterns and provenance | LlamaIndex/pgvector is an optional extra and is unavailable unless installed and configured | Selected; exact catalogue is implemented and semantic retrieval is conditionally implemented |
| Haystack with separate store | Retrieval-oriented framework | Introduces a second RAG stack without a stated capability advantage | Not selected |
| Model parametric memory only | No retrieval service | No reviewed grounding or rule provenance | Not selected |
| Rule-per-document with Oracle-side matching and PostgreSQL-side payload | Preserves atomic rule meaning and separates matching from returned guidance | Requires a curated, versioned rule corpus | Selected; represented by the knowledge-item model and retrieval interface |
| Chunked rules or hybrid BM25 retrieval | Familiar RAG techniques | Splits atomic mappings or adds complexity without a matching retrieval requirement | Not selected |

The final framework combination is a YAML-backed exact rule catalog as the default authority, with optional LlamaIndex and pgvector semantic retrieval for unmatched contextual cases. This combination is suitable for the evaluated requirements because exact mappings are deterministic and inspectable, while the optional semantic path can provide related reference patterns without replacing the exact catalog. Haystack and generic chunked-retrieval approaches were considered but are not selected [9]–[14]. The knowledge component owns rule loading, exact lookup, contextual retrieval, and rule provenance. It does not parse Oracle source, generate PostgreSQL, select a migration strategy, or decide whether generated code passes validation. This boundary prevents retrieved context from being treated as either a translation decision or a correctness verdict. At the interface level, the component returns applied or retrieved rule records, including identifiers and versions, to migration intelligence. Those records are retained in the migration artifact so the origin of a deterministic rewrite or contextual prompt input can be inspected later. Semantic retrieval is conditional on configuration and strategy needs; the exact rule catalog remains available without the optional retrieval dependencies.

The main failure modes are an incomplete or incorrect rule catalog, unavailable semantic retrieval, and a contextual rule that is relevant but insufficient for a translation. Controls include deterministic exact-key lookup, validated catalog loading, typed retrieval results, optional retrieval rather than mandatory dependency, and subsequent PostgreSQL validation. Accepted fixes are not written back into the corpus, so the current implementation remains a curated read-mostly knowledge base rather than a self-improving loop.

### D. *Validation and Executable Verification*

Validation and executable verification provide the independent evidence boundary of the migration system. The component receives a PostgreSQL candidate and assesses it using checks that do not rely on the generating strategy. Its purpose is to establish the strongest available target-side evidence, assign an explicit verification level, and return diagnostics that can support bounded repair without allowing a model or a static check to declare correctness on its own.

The final validation combination is SQLFluff for applicable pre-execution linting, Testcontainers and PostgreSQL for isolated target-side execution, PL/pgSQL and structural checks for installed code, and deterministic scenarios and invariants for behavior that can be exercised without an Oracle reference. This combination is suitable because it uses PostgreSQL installation and execution whenever the dependency context permits, while retaining static checks when executable validation is unavailable [39]–[46], [50]–[58].

The component owns candidate linting, dependency-aware installation, target-side checks, scenario execution, invariant evaluation, and typed validation results. It does not modify the candidate, select a replacement strategy, or operate the repair loop. The validation package is kept independent from agent and strategy internals so that the component judging an artifact cannot silently depend on the component that generated it. Tables XXV and XXVI summarize the static/executable validation environment and the outcome-evaluation alternatives.

TABLE XXV. STATIC VALIDATION AND EXECUTION ENVIRONMENT

| Framework or approach | Strengths considered | Limitations | Status |
|---|---|---|---|
| SQLFluff and plpgsql_check | Complementary SQL lint and PL/pgSQL semantic checks | Both checks must be maintained against supported dialects and database versions | Selected; validation code and dependencies are present |
| Guardrails valid_sql | Schema-oriented validation framework | Not a dependency or implemented validation stage in the reference repository | Not selected; architecture-only proposal |
| Testcontainers PostgreSQL environment | Reproducible real PostgreSQL validation with network-isolation support | Requires Docker and reuses one database container for a run, rather than isolating each unit in a fresh container | Selected; implemented |

TABLE XXVI. OUTCOME EVALUATION

| Framework or approach | Strengths considered | Limitations | Status |
|---|---|---|---|
| Structured execution, scenario, and invariant verdicts | Produces explicit verification levels and conservative non-pass outcomes | Does not by itself establish Oracle-to-PostgreSQL differential equivalence | Selected; implemented comparator and validation pipeline |
| Great Expectations and Soda Core differential comparison | Rich structural and aggregate differential assertions | Neither framework is an implementation dependency or demonstrated validation path | Not selected; architecture-only proposal |
| pgTAP tests through pytest | PostgreSQL-native, machine-readable tests | Not used by the reference validation implementation | Not selected |
| DeepEval GEval and DAGMetric | Advisory semantic and rule-adherence assessment | Not a dependency or implemented acceptance signal | Not selected; architecture-only proposal |

At the interface level, validation consumes a migration artifact and the dependency context produced by analysis. It emits stage-level records, diagnostics, a verification level, and a final typed outcome. Failed attempts are rolled back, while accepted artifacts remain available to later dependency groups. When required dependencies are absent, the component reports a reduced verification level rather than treating unavailable execution evidence as a pass.

Documented failure modes include syntax or installation errors, routine-level semantic errors, unavailable dependencies, Docker or database failures, and checks that are narrower than the original Oracle behavior. The corresponding controls are layered stages, typed non-pass outcomes, transactional attempt isolation, explicit missing-dependency reporting, and conservative verification labels. The absence of an Oracle runtime remains the principal limitation: this implementation cannot make a live cross-database equivalence claim.

*E. Feedback and Optimization*

Feedback and optimization control what happens when validation does not accept a candidate. The component must prevent unbounded retries while allowing a failed unit to be repaired or re-routed to a more capable eligible strategy. It also provides the evidence trail needed to understand the outcome of each attempt without making observability availability a precondition for migration.

The final framework combination is a bounded LangGraph correction flow, filesystem-based run artifacts and reports, and optional fail-open Langfuse tracing. This combination is suitable for the implemented local workflow because it makes retry state and terminal outcomes explicit without introducing external experiment tracking or a human-review service. MLflow, DSPy, and review interfaces were considered in planning but are not part of the selected runtime architecture. The component owns repair sequencing, re-routing after a failure, attempt limits, terminal outcomes, and persistence of the run evidence. It does not re-parse the source, change the exact rule catalog, or independently judge a candidate. Unlike the earlier architecture proposals, it does not pause for human approval, collect reviewer labels, or write accepted corrections back into retrieval.

At the interface level, feedback consumes validation outcomes and diagnostics, returns a repair or re-routing decision to the migration workflow, and writes attempt artifacts and reports to the run directory. Langfuse receives optional model traces through the LiteLLM callback, but authoritative report values are derived from local artifacts. This design means tracing failure reduces diagnostic visibility without invalidating a run. Tables XXVII and XXVIII summarize the correction, review, observability, and optimization options.

TABLE XXVII. CORRECTION FLOW AND REVIEW

| Framework or approach | Strengths considered | Limitations | Status |
|---|---|---|---|
| LangGraph correction flow with configurable checkpointer | Explicit bounded repair and re-routing logic | The packaged checkpoint dependency is SQLite; no PostgreSQL-backed human-review workflow is evidenced | Selected |
| Escalation ladder across repair and stronger strategies | Can use lower-cost remediation before declaring failure | Thresholds and policy choices require calibration | Selected |
| Streamlit, Gradio, or Label Studio review UI | Supports human approval, editing, and annotation | No such user-interface dependency or implementation is present | Not selected |

The principal failure modes are incorrect repair thresholds, repeated model or validation failures, unavailable tracing, and exhausted attempts. Controls include configured caps, strategy eligibility checks, explicit terminal statuses, persisted diagnostics, and fail-open observability. The residual limitation is that improvement remains manual and offline: there is no implemented human-review, experiment-registry, or learning-feedback capability.

TABLE XXVIII. OBSERVABILITY AND OPTIMIZATION OPTIONS

| Framework or approach | Strengths considered | Limitations | Status |
|---|---|---|---|
| Optional Langfuse tracing | Captures LLM traces through the LiteLLM callback and fails open when unavailable | Optional extra; deployment and endpoint availability are environment-dependent | Selected |
| MLflow experiment tracking and registry | Supports experiment comparison and model/prompt lifecycle management | Not a dependency or implemented integration | Not selected |
| DSPy optimization | Can optimize prompts from labelled feedback | Not a dependency or implemented workflow | Not selected |

## XII. CODE MIGRATION COMPONENT FIT AND TRADE-OFFS

The selected design combines framework capabilities with project-owned policies. Suitability is assessed here against capability coverage, independence of correctness evidence, integration complexity, maintainability, and operational risk. It is not a claim that these choices are optimal for every migration programme, particularly programmes with an Oracle reference environment or a production-scale human review process. Table XXIX summarizes the resulting component fit and trade-offs for the migration system.

TABLE XXIX. CODE MIGRATION COMPONENT FIT AND TRADE-OFFS

| Component | Selected framework or approach | Capability fit | Principal benefit | Main trade-off and residual risk | Evidence status |
|---|---|---|---|---|---|
| Code analysis and parsing | Project-maintained ANTLR parser and NetworkX graph planning | High for PL/SQL units and dependency-aware migration groups | Preserves source boundaries, dependency information, and parse confidence before generation | Grammar gaps, dynamic SQL, and missing source definitions can constrain analysis | Parser, graph, and regression tests are implemented |
| Migration intelligence | LangGraph, LiteLLM, deterministic routing, Ora2Pg, and rule application | High for bounded multi-strategy translation | Avoids LLM calls where deterministic translation is eligible and preserves per-attempt evidence | Thresholds and model availability affect performance; LLM output remains probabilistic | Workflow, strategy, and client modules are implemented |
| Knowledge and rules | Exact YAML catalog plus optional LlamaIndex and pgvector | High for authoritative mappings with contextual grounding | In the implemented retrieval order, exact-rule matches take precedence over similarity-ranked contextual retrieval; contextual rules retain provenance | Curated coverage may be incomplete; optional semantic retrieval adds service dependencies | Exact catalog is implemented; semantic retrieval is conditional |
| Validation and executable verification | SQLFluff, Testcontainers, PostgreSQL checks, scenarios, and invariants | Moderate to high for executable target-side evidence | Real target-database installation and execution are stronger than linting alone | No Oracle runtime prevents a differential-equivalence claim | Validation pipeline and integration tests are implemented |
| Feedback and optimization | Bounded LangGraph repair, filesystem artifacts, and optional Langfuse | Moderate for automatic recovery and diagnosis | Limits retries and retains reproducible diagnostic evidence | No human review, experiment registry, or learning loop | Repair, reporting, and optional tracing paths are implemented |

The clearest architectural fit in the evaluated implementation is obtained where a framework removes substantial engineering burden without obscuring a domain decision. ANTLR supplies grammar parsing, NetworkX supplies graph algorithms, Testcontainers manages the PostgreSQL lifecycle, and LangGraph expresses the bounded control flow. In contrast, complexity weights, eligibility policy, exact mappings, verification semantics, and status reporting remain explicit application-owned contracts.

The central trade-off is deliberate complexity. The project maintains a parser, dependency model, rule catalog, strategy profiles, validation environment, repair policy, and run artifacts. Each layer corresponds to a separate failure mode: uncertain source boundaries, invalid ordering, unnecessary model cost, ungrounded mappings, false acceptance, or irreproducible outcomes. The implementation rejects additional planned complexity where it lacks current evidence, including SQLGlot, LangChain, DSPy, MLflow, DeepEval, interactive human review, and Oracle-differential tooling.

## XIII. Validation, Human Review, Reliability, and Operational Risks

The system bases its reliability claims on visible evidence tiers rather than an aggregate quality score. A candidate is assessed through SQL linting, PostgreSQL installation, structural and PL/pgSQL checks, deterministic execution scenarios, and source-derived invariants. The resulting status and verification level must remain separate: an accepted but partially verified unit does not carry the same evidential weight as an executable unit whose dependencies were available [39]–[46], [50]–[58].

Human review is not implemented. The current tool preserves diagnostics for offline human inspection but cannot claim reviewer approval, inter-reviewer agreement, or continual improvement from approved corrections. Likewise, without an Oracle runtime it cannot demonstrate live cross-database equivalence.

Further research should measure parser coverage, routing cost and repair outcomes on representative Oracle estates, validate dependency provisioning under larger workloads, and establish an independent Oracle reference environment where governance permits. Before production use, the critical open questions are the supported Oracle dialect range, secure model-access configuration, Docker capacity under concurrency, governance of deterministic rule changes, and the process for reviewing units that terminate without full verification.

The principal migration reliability and operational risks are summarized in Table XXX.

TABLE XXX. Code Migration Reliability and Operational Risks

| Risk | Consequence | Implemented control | Remaining limitation |
|---|---|---|---|
| Unsupported, damaged, or newer Oracle syntax | Partial analysis or unsafe automation | Parse confidence, diagnostics, lexical fallback, and explicit unsupported outcomes | Coverage is limited by grammar maintenance and source quality |
| Missing or incorrect dependency information | Incorrect order or incomplete validation context | Typed graph artifacts, SCC planning, external-object reporting, and graph tests | Dynamic SQL and absent definitions remain difficult to resolve |
| Static translation pass-through or erroneous LLM output | Plausible but invalid PostgreSQL | Common validation pipeline for static and LLM outputs; bounded repair | Target-side checks cannot prove full Oracle semantic equivalence |
| Missing dependencies | Candidate cannot be executed completely | Explicit missing-dependency diagnostics and reduced verification levels | Reduced verification is weaker evidence than execution |
| Docker or PostgreSQL environment failure | An inconclusive validation result | Testcontainers lifecycle controls, typed outcomes, and persisted artifacts | Docker is required for executable validation |
| Model or optional retrieval outage | LLM-assisted migration may be unavailable | Strategy eligibility rules and deterministic path where applicable | Complex units may have no eligible alternative |
| Langfuse outage | Loss of trace visibility | Fail-open tracing; reports are derived from run artifacts | Fine-grained trace diagnosis is unavailable until recovery |
| Misconfigured repair limits | Excessive cost or premature terminal failure | Configured attempt and escalation limits with explicit terminal statuses | Thresholds require representative-corpus evaluation |

## XIV. Comparative Evaluation

### A. Comparative question and interpretation framework

The comparative question is how the three application architectures combine probabilistic GenAI components with deterministic control, and what evidence supports the suitability of those combinations for their intended tasks. The comparison is descriptive and evidence-bounded. It does not assume that a retrieval percentage, a migration acceptance rate, and an automation routing rate measure the same construct. Instead, each result is interpreted against the application's objective, its component boundary, and the strength of the available evidence [2]–[11], [56]–[58].

The comparison uses five dimensions: capability coverage, quality of the principal application outcome, reliability of the control boundary, operational efficiency, and evidence maturity. Capability coverage describes whether the selected components address the stated functional requirements. Quality is reported using the outcome metric available for each application. Reliability concerns explicit handling of invalid, unavailable, or unsafe states. Operational efficiency includes latency, model calls, tokens, and additional services where these were measured. Evidence maturity distinguishes implemented and tested behavior from prototype limitations and unmeasured production properties. The comparison uses the reported experimental results and observed behavior of the implemented systems. The chatbot retrieval experiment uses an 80-node corpus and three repeats for the storage comparison; its routing experiment uses the dataset and three repeats. The migration experiment uses 1,000 input pairs, 1,448 decomposed tasks, and seed 42, with one run per configuration. The relevant contracts, validation paths, and architectural constraints are treated as implementation observations rather than as independent comparative references. The Automation evidence base includes a labelled triage campaign measured over repeated runs (classification and routing accuracy plus escalation rate) and programmatic runtime-guardrail evaluation (precision/recall on a labelled violation set), and it records evidence maturity explicitly by distinguishing implemented behavior from specified-but-unintegrated evaluation/observability surfaces.

### B. Evaluation evidence by application

Table XXXI summarizes the principal measured results. Percentages are retained with their denominators or experimental scope where the source provides them. The results should not be aggregated into a single cross-application score. The chatbot results support retaining hybrid retrieval and the custom PostgreSQL integration, while they do not support enabling the cross-encoder reranker by default. They also show that a system can have strong retrieval performance and still have a materially weaker routing boundary: the retrieval metrics and routing correctness measure different components and should not be conflated. The migration results support a complementary conclusion: deterministic rules are highly effective on the subset they cover, whereas the combined strategy is preferable when the objective includes broad task coverage. The agent's advantage over the LLM-only baseline is measurable but modest in token usage, so its main benefit is the joint quality-and-coverage profile rather than cost minimization alone.

### C. Comparative fit of architectural patterns

Table XXXII compares the architectural patterns rather than treating framework names as interchangeable products. The entries marked as implemented are supported by repository behavior; for Automation, several control-boundary and durability properties are implemented and measured, while specific monitoring backends and the human-feedback-to-dataset loop remain unintegrated.

TABLE XXXI. PRINCIPAL MEASURED RESULTS ACROSS THE THREE APPLICATIONS

| Application | Compared alternatives or baseline | Principal measured result | Interpretation | Evidence limitation |
|---|---|---|---|---|
| Chatbot development assistant | Hybrid dense plus lexical retrieval with RRF versus BM25-only | Recall@8: 94.7% versus 15.8%; R-Precision: 81.6% versus 13.2%; MRR: 90.3% versus 16.7%; Precision@k: 46.4% versus 66.7% | On the evaluated corpus, hybrid retrieval produced substantially higher recall and ranking metrics than the BM25-only baseline for the evaluated documentation set, at the cost of lower Precision@k than the sparse-only baseline | retrieval@v1 was not execution-filtered; absolute metrics require caution |
| Chatbot development assistant | Cross-encoder reranker enabled versus disabled | MRR: 89.7% versus 90.3%; R-Precision: 73.7% versus 81.6%; no Recall@8 change; approximately +213 ms/query with reranking | The reranker added latency without improving the measured retrieval outcomes and is disabled by default | Results are corpus- and model-specific; no larger-scale latency study is reported |
| Chatbot development assistant | Current PostgreSQL store versus LlamaIndex PGVectorStore | Recall@8: 94.7% versus 92.1%; R-Precision: 81.6% versus 76.3%; MRR: 90.3% versus 91.9%; mean latency: 459 versus 443 ms over 108 calls | The custom store leads on Recall and R-Precision and preserves project-specific metadata as first-class fields; the stock store shows a marginally higher MRR and a 16-ms lower mean latency in the evaluated runs; the available experiment does not establish whether that latency difference is practically or statistically meaningful. | The comparison covers one corpus, one embedding model, and one top_k value |
| Chatbot development assistant | Current router against expected tool choice and arguments | Tool-choice correctness: 85.6%; argument correctness: 87.2% over three repeats | The routing path achieved 85.6% tool-choice correctness and 87.2% argument correctness in the evaluated runs, with repeatable failure cases remaining | This is not a framework A/B comparison and does not establish general routing quality |
| Automation email/request triage | Classification: shipped intent descriptions vs rewritten descriptions, on a small deployment (frozen 40-case set; 5 repeats) | Accuracy: 0.897 → 0.943 (+4.6 pp) | Description/prompt rewriting improved measured classification on the weakest evaluated deployment but not consistently across the other deployments, suggesting compensation for small-model boundary weakness rather than a universal prompt improvement | 40-case labelled set is small and self-authored (self-authored human-labelled evidence without independent annotation); result is deployment-specific; not validated on real traffic |
| Automation email/request triage | Classification: shipped intent descriptions vs rewritten descriptions, on a mid-tier deployment (same frozen 40-case set) | Accuracy: 0.943 → 0.909 (−3.4 pp) | The same rewrite that helped the smallest deployment regresses on a different deployment, reinforcing that tuning results are not portable across model classes | Same limitations as above; only a single frozen set is reported |
| Automation email/request triage | Classification: shipped intent descriptions vs rewritten descriptions, on a larger deployment (same frozen 40-case set; 5/5 repeats stable) | Accuracy: 0.971 → 0.971 (0.0 pp) | On the largest evaluated deployment, the rewrite produced no measured accuracy change, which is consistent with—but does not by itself prove—the interpretation that the rewrite compensated for weaknesses of the smaller deployment. | Same dataset/label limitations; does not establish performance under distribution shift |
| Automation email/request triage | Operational efficiency: three deployment classes for the same 40-case repeat | Mean per-call latency ≈ 1385 / 6016 / 1797 ms and tokens ≈ 37k / 55k / 73k per 40-case repeat (small/mid/large) | The evaluated deployments exhibited materially different latency, token usage, and classification behavior; the observed latency and token-use differences are large enough to merit consideration as architectural constraints in this evaluated configuration | The reported measurements are specific to one gateway path, one workload shape, and one frozen set; not a production throughput study |
| Automation email/request triage | Live triage campaign: LLM-based decision layer vs a keyword baseline control arm (40 labelled cases; 5 repeats; 2 arms; one small deployment) | Classification 89.7% (spread 2.9 pp), routing 98.7% (spread 3.3 pp), escalation rate 72.5% (spread 0.0 pp) | The measured outcome metrics are interpretable as application-specific quality signals (classification/routing) plus workload signal (escalation), not as a construct comparable to retrieval recall or migration acceptance | Labels are self-authored; the keyword baseline scores 0 on natural prose and is a floor rather than a competitive alternative; escalation is not an accuracy metric |
| Automation email/request triage | Runtime guardrail effectiveness on labelled violation set (pattern-based guardrails) | Precision 0.733, recall 0.688 on 29 labelled drafts; false-block rate 0.308; reason-code accuracy 1.000 | Guardrails provide measurable live blocking but exhibit two distinct failure modes (surface-form evasion vs numeric-shape collision), so reporting must preserve precision and recall rather than a blended score | The set is adversarially constructed to probe failures; results characterize failure-mode shape, not production base rates |
| Oracle-to-PostgreSQL migration | Task-decomposed strategy selection versus LLM-only and rules-only baselines | Accepted tasks: 91.9% versus 90.5% versus 14.2%; high structural similarity (at least 0.9): 426 versus 379 versus 142 | In the evaluated run, combining deterministic rules with LLM strategies produced higher accepted-task coverage and structural agreement than the evaluated LLM-only baseline. | Structural similarity is not behavioral equivalence; human references were not independently audited |
| Oracle-to-PostgreSQL migration | Same three approaches, target-side validation | Installed or executed tasks: 284 for the agent, 275 for LLM-only, and 6 for rules-only | The agent produced more artifacts reaching PostgreSQL evidence than both evaluated baselines, with the largest difference relative to the rules-only baseline. | Most tasks reached only static checks, and missing dependencies limited execution |
| Oracle-to-PostgreSQL migration | Agent pipeline versus LLM-only baseline | 10,020,728 versus 10,332,904 tokens (-3.0%); 1,647 versus 1,770 calls (-6.9%); wall time 54.9 versus 54.1 minutes with 20 workers | Strategy selection was associated with 3.0% fewer tokens and 6.9% fewer model calls than the evaluated LLM-only baseline in the reported run, while producing higher measured structural outcomes | Each configuration was run once; the stronger model was not evaluated as an all-task baseline |

TABLE XXXII. CROSS-APPLICATION COMPARISON OF ARCHITECTURAL PATTERNS

| Evaluation dimension | Chatbot development assistant | Automation email/request triage | Oracle-to-PostgreSQL migration |
|---|---|---|---|
| Principal probabilistic task | Grounded answer generation, selective retrieval, and tool-use proposal | Intent classification, department routing proposal, and reply drafting | Contextual code translation and bounded repair |
| Deterministic decision boundary | Typed orchestration state, citation checks, tool schemas, read/write declarations, and approval binding | Runtime draft guardrails (pass/block/revise), deterministic routing lookup, versioned authorization rules with allow/block/needs-approval, durable ledger/state, idempotency at run and action levels, and human approval gating side effects | Parser and dependency graph, deterministic strategy eligibility, PostgreSQL validation, and terminal outcomes |
| Selected control-flow style | Synchronous LangGraph workflow with PostgreSQL checkpoints | Asynchronous durable workflow with explicit pause–resume for HITL: Temporal durability combined with a graph-style control-flow representation; composition in the evaluated version pairing required a topology/predicate declaration shared across executors because the tested LangGraph graph did not execute successfully inside Temporal's workflow sandbox | LangGraph correction flow with a run-level DAG executor |
| Grounding or knowledge strategy | Hybrid dense/lexical retrieval with RRF; citation metadata preserved | Versioned routing-rule snapshot (intent → department/description) plus external rule store for authorization policy; no retrieval-augmented knowledge base is a required dependency for triage correctness in the reference build | Exact YAML rule catalog, with optional semantic retrieval for contextual cases |
| Independent correctness evidence | Retrieval metrics, citation checks, tool-contract tests, and replay/approval tests | Labelled triage campaign metrics (classification/routing accuracy; escalation rate treated as workload), runtime-guardrail precision/recall on a labelled violation set, and execution-tier evidence for durable pause–resume and fail-closed/idempotent side-effect control | PostgreSQL installation/execution, scenarios, invariants, and typed verification levels |
| Main measured strength | Retrieval recall and ranking improve strongly over sparse-only retrieval | Strong deterministic control boundary: fail-closed authorization under dependency outage, replay-safe idempotency, durable pause–resume across worker replacement; routing step is deterministic (token cost 0) and guardrails are measured as precision/recall rather than a single blended score | Higher accepted-task coverage and more high-similarity outputs than the two evaluated baselines in the reported run |
| Main measured weakness | Routing and argument correctness remain below full reliability | Outcome-quality evidence is limited by small, self-authored labelled sets; classification "improvements" are deployment-dependent; measured guardrails show both surface-form evasion (misses) and numeric-shape collision (false blocks); observability backends and the human-review→dataset feedback loop are specified but not integrated | Repair and escalation consume substantial tokens; no Oracle runtime establishes equivalence |
| Operational maturity | Prototype with live-tool credentials and production-scale corpus still limited | Reference prototype demonstrates end-to-end gating and durability, but monitoring/export infrastructure is incomplete (no tracing backend wired; feedback loop not implemented) | Local reference implementation; executable validation requires Docker and the required dependency context |

Across the evaluated applications, the clearest evidence of framework fit appears where the framework capability maps directly to a mechanically testable requirement. LangGraph and checkpointing make chatbot state transitions inspectable; the migration workflow makes repair and terminal outcomes explicit. In Automation, the durable-execution model makes pause–resume and replay safety mechanically testable, and the authorization/guardrail gates make "LLM proposes; deterministic decides" enforceable at the side-effect boundary rather than as a convention [9]–[11], [18]–[23], [38]–[49], [50]–[55].

The applications also demonstrate different meanings of "grounding." In the chatbot, grounding means retrieving relevant documentation and preserving citation provenance. In migration, grounding means applying an authoritative exact rule before using similarity-based contextual guidance and then validating the generated target artifact. In Automation, "grounding" is primarily policy grounding: decisions are grounded in versioned routing and authorization rules and in deterministic guardrail verdicts rather than in retrieval of external knowledge. In all three systems, retrieval or generation is subordinate to a deterministic evidence boundary: a retrieved chunk does not by itself prove an answer, a retrieved migration rule does not by itself prove semantic correctness, and a drafted reply does not by itself authorize an external side effect.

The comparison yields four findings, summarized in Table XXXIII.

### D. Comparative conclusions

For the chatbot development assistant, the evidence supports retaining hybrid retrieval with the custom PostgreSQL store, disabling the reranker by default, and treating routing quality as an open improvement area. The architecture is consistent with the evaluated requirements of a documentation-grounded assistant whose primary risks are unsupported answers, stale or unavailable tools, and uncontrolled side effects, but its current retrieval results should not be generalized beyond the evaluated corpus.

For the Oracle-to-PostgreSQL migration system, the evidence supports a mixed strategy architecture. Rules alone provide accepted deterministic translations for a narrow subset covered by the implemented rule set; the evaluated LLM-based strategies expanded accepted-task coverage relative to the rules-only baseline; and deterministic routing with validation combines these strengths. The measured results show higher structural agreement and validation reach than the evaluated LLM-only baseline; however, the single-run design does not establish statistical significance or stability across repeated runs, and the small token reduction does not justify a claim of broad cost efficiency. Neither structural agreement nor target-side validation establishes Oracle-to-PostgreSQL semantic equivalence.

TABLE XXXIII. CROSS-APPLICATION FINDINGS AND IMPLICATIONS

| Finding | Comparative evidence | Interpretation and implication |
|---|---|---|
| Component specialization improves evidence quality when boundaries are explicit. | The chatbot separates retrieval, tool execution, and answer generation; the migration system separates parsing, strategy selection, generation, and validation. The Automation system separates probabilistic proposal (classification/drafting) from deterministic validation and authorization, and sequences them via an explicit orchestration layer. Across repositories, these boundaries are realized through typed contracts and targeted tests/fault injection. | Independent boundaries expose distinct failures and make them locally diagnosable: irrelevant retrieval and invalid tool arguments (chatbot), unsupported syntax and failed target-side installation (migration), and in Automation, guardrail-triggered revise/block outcomes, fail-closed authorization under dependency outage, and idempotency/replay defects that are detectable at the gate rather than after an external side effect. |
| Deterministic controls constrain consequences but do not eliminate probabilistic quality risk. | Chatbot tool validation prevents malformed or unauthorized calls, but measured tool-choice correctness remains 85.6%. Migration validation rejects or downgrades unsupported candidates, but does not establish Oracle behavioral equivalence. Automation enforces “LLM proposes; deterministic decides”: runtime guardrails and a deterministic authorization gate prevent side effects without passing explicit gates, but measured classification quality is deployment-dependent and runtime guardrails have nonzero false blocks and misses (precision 0.733, recall 0.688 on a labelled adversarial set). | Deterministic layers improve auditability and safety by limiting what can happen, but they do not guarantee that probabilistic outputs are correct or useful; outcome quality remains an empirical property that must be evaluated on representative data and may vary with model deployment. |
| The most suitable architecture is not necessarily the one with the lowest model usage. | The migration agent used 3.0% fewer tokens than the LLM-only baseline while increasing accepted tasks and high-similarity outputs. Hybrid chatbot retrieval improved recall and ranking metrics but had lower Precision@k than BM25-only retrieval. In Automation, routing resolution is implemented as deterministic lookup (0 model calls), yet overall robustness still depends on the chosen model for classification/drafting, with measured latency/token trade-offs across deployments. | Suitability should be assessed against the application’s primary objective and the full quality–cost–risk trade-off rather than a single efficiency metric; reducing model usage in one step does not remove probabilistic error elsewhere, and stronger models may be justified by boundary robustness even at higher cost. |
| Evidence maturity limits cross-application generalization. | The chatbot retrieval and storage comparisons used repeated measurements on a small corpus, whereas the migration comparison used a larger sample but one run per configuration. Automation contributes measured campaign and guardrail results, but some evaluation/observability surfaces are specified yet unintegrated (e.g., tracing backends and the human-review→dataset feedback loop), and some outcome figures rely on small self-authored labelled sets. | The findings support conditional recommendations tied to workload, control requirements, and evidence quality; they do not establish universal framework superiority. In particular, Automation’s control-boundary evidence is stronger than its outcome-quality evidence, limiting how far its quantitative figures can be generalized beyond the measured campaigns. |

For Automation, the evidence supports the suitability of the control boundary pattern—LLM proposals gated by deterministic validation and authorization—more strongly than it supports any broad claim about outcome quality. A labelled triage campaign reports classification and routing accuracy with an explicit escalation rate as a workload signal, and a programmatic runtime-guardrail evaluation reports precision/recall on a labelled violation set. However, these results are bounded by small, self-authored and/or adversarially constructed datasets, and measured classification behavior varies by model deployment, indicating that prompt- or description-level “improvements” are not portable across models. In addition, several evaluation-maturity elements described in the method (notably observability backends and the human-review→dataset feedback loop) are specified but not integrated in the reference build, limiting the strength of longitudinal or production-like conclusions.

Taken together, the results favor a layered architecture in which probabilistic components perform tasks requiring language or code-generation capability, while deterministic components own routing constraints, validation, authorization, persistence, and evidence recording. The reported framework choices should be interpreted as conditionally suitable for the evaluated component contracts, workloads, framework versions, and deployment assumptions rather than as generally superior alternatives. The principal research gap is not the absence of additional frameworks; it is the need for larger, repeated, independently labelled, and application-specific evaluations—plus end-to-end wiring of the monitoring and feedback surfaces—that can test whether the implemented controls and measured quality signals continue to hold under realistic workload, model-deployment, and dependency-failure conditions [1]–[11], [26]–[40], [56]–[58].

## XV. LIMITATIONS AND OPEN QUESTIONS

The findings are constrained by differences in application maturity, experimental design, and available validation evidence. The three applications were not evaluated through one uniform benchmark because their outputs and failure consequences are not commensurate. Retrieval relevance, workflow classification, side-effect safety, structural code similarity, and target-database execution measure different properties. The study therefore supports conditional architectural conclusions rather than a numerical ranking of the applications or framework families.

Several validity threats follow from these limitations. Construct validity is restricted where a metric measures only part of the intended outcome, such as structural similarity without behavioral equivalence, retrieval correctness without answer correctness, or triage accuracy without measuring end-to-end user impact. Internal validity is restricted where configurations were not repeated sufficiently to characterize variance (e.g., single-run migration comparisons) and where model deployment is a confounded factor (e.g., Automation’s deployment-conditioned prompt effects). External validity is restricted by small or authored datasets, partial live-system access, and the absence of sustained production workloads. Reproducibility also depends on retaining model identifiers, configuration versions, corpus state, rule versions, and validation-environment details, and on recording evaluation in a form that can be re-run [38], [49]–[52], [56], [57].

The analysis further assumes that explicit component boundaries improve maintainability and auditability. The implementations provide evidence that these boundaries expose failures and support targeted tests, but the study does not quantify their long-term maintenance cost. Additional contracts and specialized frameworks may reduce behavioral ambiguity while increasing dependency management, deployment effort, and operational ownership. This trade-off remains application-dependent. Table XXXIV summarizes the principal limitations and the corresponding open research questions.

TABLE XXXIV. PRINCIPAL LIMITATIONS AND OPEN RESEARCH QUESTIONS

| Area | Current limitation | Open question |
|---|---|---|
| Cross-application comparison | Evaluation datasets, repetitions, baselines, and outcome measures differ across applications, and evidence strength differs by axis (e.g., retrieval metrics vs. human-labelled triage vs. execution-tier validation). | Which common reporting protocol (score record schema, per-axis spread, explicit baselines, evidence-tier annotation) can improve comparability without collapsing application-specific outcomes into an invalid composite score? |
| Chatbot corpus and retrieval | Retrieval experiments use a small technical corpus, and the labelled retrieval set was not execution-filtered. | Do the observed hybrid-retrieval and storage results persist with a larger, continuously changing corpus and independently reviewed labels? |
| Chatbot answer quality | Retrieval, citation, and routing properties are measurable, but free-form answer helpfulness lacks an independent automatic oracle. | Which human evaluation design can assess correctness and usefulness without conflating source quality, retrieval quality, and generation quality? |
| Chatbot integrations and security | Live credentials, per-user authorization, production data residency, and sustained concurrency are not comprehensively evaluated. | Can the approval, redaction, and tool-isolation controls preserve their properties under realistic identity models and production traffic? |
| Automation evaluation (outcome quality) | Automation contributes quantitative results (classification/routing metrics, escalation rate; runtime guardrail precision/recall), but they are bounded by small labelled sets, including self-authored triage labels without independent annotation and an adversarially constructed guardrail set that characterizes failure modes rather than production base rates. | How do outcome-quality and workload metrics (classification/routing, escalation rate, revise-loop frequency, approval rate) behave on larger, independently labelled, real-traffic-derived datasets across repeated runs? |
| Automation evaluation (model dependence) | Automation classification behavior varies by model deployment: a description rewrite improved the weakest deployment, regressed a mid-tier deployment, and was inert on the strongest, indicating deployment-conditioned results rather than portable prompt improvements. | What is the minimal model deployment (cost/latency) that achieves stable boundary behavior for the chosen intent vocabulary, and how should deployment changes be treated as experimental factors in evaluation reporting? |
| Automation operations and monitoring maturity | Durable pause–resume and gating are implemented, but several operational surfaces remain incomplete: no integrated tracing backend, no implemented human-review→dataset feedback loop, and inbound DLQ recovery is not closed-looped by a consumer/process. | Which operational surfaces (approval notification UI/channel, DLQ triage workflow, policy administration, tracing/alerting backend, human-feedback ingestion) are required to operate the workflow end-to-end outside the orchestration engine under realistic failure and workload conditions? |
| Migration equivalence | Validation is primarily target-side because no Oracle execution environment is available. Structural agreement and PostgreSQL execution do not prove behavioral equivalence. | How do accepted artifacts compare with Oracle behavior under differential tests on representative schemas and data? |
| Migration dataset and repetitions | The comparative migration configurations were run once, all three configurations share the same task-decomposition stage so the effect of decomposition itself is not isolated, many source fragments lack dependencies required for installation or execution, and the stronger-model tier was applied only to the hardest fragments rather than evaluated as an all-task baseline. | Are the observed acceptance, similarity, and resource differences stable across repeated runs and dependency-complete enterprise corpora, and does the stronger model justify its cost when evaluated fairly on the full task set? |
| Migration strategy policy | Repairs and stronger-model escalation consume substantial resources, while some LLM calls replace exact rules-based outputs. | Which deterministic stopping and retention policies maximize validated coverage per model call? |
| Framework evolution | Dependency compatibility and process-level interactions were evaluated for specific versions and deployment conditions; some composition issues are specifically process-wide (e.g., sandbox/import-hook interactions). | Which compatibility findings persist after framework, model, gateway, or runtime upgrades, and what regression tests should be mandatory at upgrade time to detect breaking cross-framework interactions? |

## CONCLUSIONS

This study compared framework suitability at the component level across three enterprise GenAI applications with different execution and correctness requirements: a documentation-grounded development assistant, an email/request automation workflow, and an Oracle-to-PostgreSQL migration system. The results show that framework suitability cannot be reduced to a single application-wide ranking. The analysis indicates the strongest architectural fit when each framework is assigned a bounded responsibility and probabilistic model outputs are separated from deterministic control over routing, authorization, persistence, validation, and terminal outcomes.

For the chatbot, hybrid dense–lexical retrieval substantially improved recall and ranking quality over the sparse-only baseline on the evaluated corpus, while routing and tool-argument correctness remained a separate reliability limitation. The evaluated cross-encoder reranker did not improve retrieval outcomes sufficiently to justify its added latency. For Automation, the strongest evidence concerned the deterministic execution boundary: durable

pause–resume behavior, fail-closed authorization, replay-safe idempotency, and runtime guardrails were directly exercised. At the same time, classification results varied across model deployments, indicating that prompt-level improvements should not be assumed to transfer between models. For Oracle-to-PostgreSQL migration, the evaluated combination of deterministic rules, complexity-aware strategy selection, LLM-assisted translation, and independent PostgreSQL validation produced higher accepted-task coverage and target-side validation reach than the evaluated rules-only and LLM-only baselines.

Across all three evaluated applications, the findings support a common architectural principle: probabilistic model outputs should remain proposals when externally observable or executable consequences require deterministic authorization or validation. The evaluated implementations also show that framework composition can introduce costs that are not visible when individual technologies are evaluated independently, including dependency constraints, process-level incompatibilities, additional operational services, and incomplete observability paths. Consequently, framework selection should be based on explicit component contracts, failure semantics, evidence requirements, and deployment constraints rather than on feature breadth alone.

The reported results remain bounded by prototype-scale datasets, application-specific evaluation procedures, limited repeated measurements for some experiments, incomplete production observability, and the absence of an Oracle reference runtime for differential migration validation. Future work should therefore evaluate the architectures on larger independently labelled datasets and production-like workloads, repeat migration experiments across model and configuration variants, complete end-to-end monitoring and human-feedback pipelines, and investigate whether the observed framework-composition constraints persist across framework and runtime versions. These extensions are necessary before the reported component-level findings can be generalized to broader enterprise deployments.

## Declaration on Generative AI

During the preparation of this work, the authors used ChatGPT (OpenAI) to check grammar and spelling. After using this tool, the authors reviewed and edited the output as needed and take full responsibility for the content of the publication.